\documentclass[journal=jctc,manuscript=article]{achemso}

\usepackage[version=4]{mhchem} 
\usepackage{booktabs}
\usepackage{color}
\usepackage{multirow}
\usepackage{longtable}
\usepackage{graphicx}
\usepackage{siunitx}
\usepackage{lscape}
\usepackage{ltablex}
\usepackage{caption}
\usepackage{subcaption}
\usepackage{cleveref}
\usepackage{xr}
\usepackage[final]{pdfpages}

\makeatletter
\newcommand*{\addFileDependency}[1]{
  \typeout{(#1)}
  \@addtofilelist{#1}
  \IfFileExists{#1}{}{\typeout{No file #1.}}
}
\makeatother

\newcommand*{\myexternaldocument}[1]{%
    \externaldocument{#1}%
    \addFileDependency{#1.tex}%
    \addFileDependency{#1.aux}%
}

\myexternaldocument{SI}
\author{Jacopo Lupi}
\affiliation{Scuola Normale Superiore, Piazza dei Cavalieri 7, I-56126 Pisa, Italy}
\altaffiliation{J.L and S.A. contributed equally to this work.}
\author{Silvia Alessandrini}
\affiliation{Scuola Normale Superiore, Piazza dei Cavalieri 7, I-56126 Pisa, Italy}
\alsoaffiliation{Dipartimento di Chimica ``Giacomo Ciamician'', Universit\`{a} di Bologna, Via F. Selmi 2, I-40126 Bologna, Italy.}
\altaffiliation{J.L and S.A. contributed equally to this work.}
\author{Cristina Puzzarini}
\affiliation{Dipartimento di Chimica ``Giacomo Ciamician'', Universit\`{a} di Bologna, Via F. Selmi 2, I-40126 Bologna, Italy.}
\email{cristina.puzzarini@unibo.it}
\author{Vincenzo Barone}
\affiliation{Scuola Normale Superiore, Piazza dei Cavalieri 7, I-56126 Pisa, Italy}
\email{vincenzo.barone@sns.it}
\title[ChS-F12 composite schemes]
  {The junChS and junChS-F12 models: parameter-free efficient yet accurate composite schemes for energies and structures of non-covalent complexes}

\abbreviations{}

\begin{document}


%




\begin{abstract}
A recently developed model chemistry (denoted as junChS [Alessandrini et al J. Chem. Theory Comput. \textbf{2020}, 16, 988-1006]) has been extended to the employment of explicitly-correlated (F12) methods. This led us to propose a family of effective, reliable and parameter-free schemes for the computation of accurate interaction energies of molecular complexes ruled by non-covalent interactions. A thorough benchmark based on a wide range of interactions showed that the so-called junChS-F12 model, which employs cost-effective revDSD-PBEP86-D3(BJ) reference geometries, has an improved performance with respect to its conventional counterpart and outperforms well-known model chemistries. Without employing any empirical parameter and at an affordable computational cost, junChS-F12 reaches sub-chemical accuracy. 
Accurate characterizations of molecular complexes are usually limited to energetics. To take a step forward, the conventional and F12 composite schemes developed for interaction energies have been extended to structural determinations. A benchmark study demonstrated that 
the most effective option is to add MP2-F12 core-valence correlation corrections to fc-CCSD(T)-F12/jun-cc-pVTZ geometries without the need of recovering the basis set superposition error and the  extrapolation to the complete basis set. 
\end{abstract}

\section{Introduction}

The quantitative description of non-covalent interactions is of paramount relevance in widely different fields ranging from biochemistry to nanoscience, going through chemical reactivity in the Earth atmosphere and in the interstellar space. \cite{francisco2002,Klemperer10584,riley2011,tang2013,puzzarini_jpc2020,sastry2014} The reason lies in the fine tuning of geometric and energetic properties issuing from the formation of non-covalent bonds, which is due to the synergistic role played by electrostatic, induction and dispersion contributions. Concerning biological model systems, non-covalent interactions are --for example-– at the basis of DNA bases pairing and complexation with solvent molecules. Moving to reactivity, in the fields of astrochemistry and combustion chemistry, pre-reactive complexes are crucial points, whose properties are ruled by weak interactions. 
Furthermore, loose transition states also belong to the class of non-covalently bonded compounds. 

Aiming at the definition of a general protocol able to accurately characterize non-covalent complexes at a non-prohibitive computational cost, the focus of this work is mainly on the interaction energy at the equilibrium geometry for medium-sized molecular systems containing up to about 20 atoms belonging to the first three rows of the periodic table. For inter-molecular complexes that do not show strong static correlation effects, the most successful strategy is based on the coupled cluster (CC) ansatz. In particular, the CC model that accounts for the full treatment of single and double excitations and the perturbative inclusion of triple excitations, CCSD(T),\cite{raghavachari1989fifth} has become the so-called ‘gold standard’ in contemporary computational chemistry. 
As a matter of fact, a fortuitous but quite general error compensation makes CCSD(T) performances often better than those delivered by the model including the full treatment of triple excitations (CCSDT)\cite{ccsdt1}, so that improvement beyond CCSD(T) results requires, together with full treatment of triple excitations, at least perturbative inclusion of quadruple excitations (CCSDT(Q)).\cite{1.2121589} 

In order to establish a protocol able to combine accuracy and efficiency, the additivity approximation is adopted and leads to the definition of composite schemes, which aim at minimizing the errors due to ($i$) the truncation of the basis set and ($ii$) the $N$-electron error associated to the CCSD(T) method. While CCSD(T) can be considered affected by a small $N$-electron error, the complete basis set (CBS) limit can be accurately estimated by means of extrapolation procedures based on hierarchies of basis sets.\cite{dbh24_f12,Helgaker97} These being computationally demanding already for medium-sized systems, the frozen-core (fc) approximation is usually employed with an additive incorporation of the core-valence (CV) correlation effects. Although CV contributions are usually small for interaction energies of molecules containing only second-row atoms, this is not the case for geometries and, when third-row atoms are present, also for energies.\cite{chemistry21} 

Focusing on molecular complexes, the accurate description of non-covalent interactions requires the presence of diffuse functions in the basis set (basis sets denoted as ``augmented'') because these interactions are particularly sensitive to the tails of the wave functions of the partners. In the last decade, a number of systematic investigations has shown that CCSD(T) computations in conjunction with ``augmented” triple- and quadruple-zeta basis sets and followed by extrapolation to the CBS limit provide, in most cases, highly accurate results.\cite{Karton2021,atlas1,atlas3} However, composite schemes entirely based on CCSD(T) calculations are computationally expensive and not affordable for larger systems, such as those of biological or prebiotic relevance. Among the approaches proposed to reduce the computational cost, we have introduced the so-called ‘cheap’ scheme (hereafter ChS) that, on top of fc-CCSD(T) calculations in conjunction with a triple-zeta quality basis set, includes the contributions due to the extrapolation to the CBS limit and CV effects using second-order Møller-Plesset perturbation theory (MP2).\cite{moller1934note} This scheme, whose denomination is due to the strong reduction of the computational cost with respect to an approach entirely based on CC computations, is well tested for semi-rigid and rather flexible systems,\cite{puzzarini2011extending,C3CP52347K,barone_2013,puzzarini2014accurate,py-nh3,oberchain} and has been recently extended to incorporate partially augmented basis sets (junChS) in order to treat molecular complexes\cite{alessandrini2020}. 

To extend the range of applicability of the ChS approach, one can exploit the fact that the extrapolation to the CBS limit can benefit from explicitly-correlated methods and resolution of identity, as well-illustrated in the literature.\cite{f12_chemrev,dbh24_f12} The rate-determining step of the junChS model is the CCSD(T) energy evaluation in conjunction with the jun-cc-pVTZ basis set.
Since the use of a double-zeta quality basis set fairly reduces the accuracy of conventional approaches, a possible solution is offered by the CCSD(T)-F12 method.\cite{knizia2009_054104} In this approach, the CCSD part benefits of the F12 implementation, thus rapidly converging to the CBS limit,\cite{Kesharwani2018} with the computational bottleneck represented by the conventional evaluation of the (T) contribution. These considerations prompted us to develop a “F12 version” of the junChS model (junChS-F12) that, on top of CCSD(T)-F12 calculations, employs MP2-F12 \cite{mp2-f12} for the extrapolation to the CBS limit. In this work, we investigate whether explicitly-correlated F12 approaches can improve the reliability and/or reduce the computational cost of the junChS model by reducing basis-set dimensions. In this regard, both basis sets specifically developed for explicitly-correlated computations (the cc-pV$n$Z-F12 and aug-cc-pV$n$Z-F12 families with $n$=2-4) \cite{peterson2008_084102} and traditional basis sets (the jun-cc-pV$n$Z family with $n$=3,4) \cite{Dunning-JCP1989_cc-pVxZ,SummerBasisSets} will be investigated in conjunction with the F12 methods. 

While accurate characterizations of molecular complexes are usually limited to their energetics, structural information is likewise important, because the energetics little depends on the structure only when reliable reference geometries are employed.\cite{Spackman2016,Karton2021b} In this respect, benchmark studies are missing or very limited. A significant example is provided by model systems to study biological interactions, where the geometry has sometimes a huge impact on the type and the strength of interactions established. Another example is offered by the investigation of reactive potential energy surfaces, which are often characterized by crucial stationary points (either intermediates or transition states) that are molecular complexes.\cite{barone2021} However, the most compelling example is the spectroscopic characterization of molecular complexes. Spectroscopic techniques are very powerful, also because they are noninvasive. The interpretation of all spectroscopic studies requires preliminary accurate structural evaluations because there is a strong relationship between the experimental outcome and the electronic structure of the system under investigation.\cite{chemrev19}
Since the accurate determination of equilibrium geometries of non-covalent complexes is a quite unexplored field, in this study, we have decided to also investigate the performance of conventional and explicitly-correlated composite schemes for this purpose. While the conventional ChS scheme is well tested for semi-rigid and flexible molecules,\cite{puzzarini2011extending,C3CP52347K,barone_2013,puzzarini2014accurate,puzzarini2018diving} its use in structural characterization of molecular complexes is limited to a few studies.\cite{py-nh3,oberchain,cyclopentene} A more systematic investigation is here carried out together with the extension of different composite schemes to the explicitly-correlated counterparts.

To conclude, we can summarize the goals of this study as follows:
\begin{itemize}
    \item Definition and validation of explicitly-correlated ChS schemes for energy evaluation;
    \item Definition and validation of conventional and explicitly-correlated composite schemes for structural determination;
    \item Definition of the junChS and junChS-F12 databases. 
\end{itemize}
The manuscript is organized as follows. In the next section, the methodology is described in detail. In the following section, the performance of the explicitly-correlated composite schemes is discussed for interaction energy first and for structural determination then. Finally, the junChS and junChS-F12 databases are introduced and analyzed.

\section{Methodology}

In the following, the computational strategies employed in our work are thoroughly described. We start by introducing the computational details and the set of inter-molecular complexes selected. Then, the ChS schemes based on F12 explicitly-correlated methods are presented. This requires, first of all, the introduction of a sort of glossary to be used along the manuscript. Finally, the application of the conventional and F12 ChS schemes to structural determinations is described.

\subsection{Computational Details}

The Gaussian \cite{g16} and Molpro \cite{Molpro,Molpro1} suites of programs have been employed for conventional and explicitly-correlated computations, respectively. Density fitting and resolution of identity approximations have been used throughout in F12 calculations. In all F12 calculations, the Slater-type geminal exponent has been set to $\gamma=\SI{1.0}{\per\bohr}$.

Concerning CCSD(T)-F12 computations, the F12b approximation\cite{f12b} has been used in combination with the extrapolation to the CBS limit, while the F12a variant\cite{f12a} has been employed in the other cases. For MP2-F12 computations, the default approximation (3C with EBC) has always been employed.\cite{mp2-f12} As already noted, in the CCSD(T)-F12 method, the perturbative treatment of triples is the same as in the conventional CCSD(T) counterpart. Indeed, a full F12 correction to the term accounting for triple excitations is not yet available in commercial codes. Although this contribution can be estimated from the difference between MP2 and MP2-F12 energies or pair contributions,\cite{knizia2009,kallay2021} we preferred to avoid any empirical term in the ChS-F12 family of composite schemes. 

As mentioned in the Introduction, specific  (aug-)cc-pV$n$Z-F12 basis sets as well as standard aug-cc-pV$n$Z and ``seasonal" basis sets have been employed in combination with F12 calculations.
Whenever aug-cc-pV$n$Z or cc-pV$n$Z-F12 orbital basis sets are used, suitable density fitting (DF) basis and resolution of the identity (RI) basis sets are automatically chosen by the Molpro program.\cite{Molpro} In all the other cases, these basis sets should be carefully selected. The general recipe adopted in this paper is to choose DF and RI basis sets corresponding to the augmented versions of the orbital ones.
It is worthwhile recalling that the generic cc-pV\textit{n}Z-F12 basis set actually corresponds to the may-cc-pV(\textit{n}+1)Z set, where the highest angular momentum functions are neglected for all atoms and an additional set of \textit{p}-functions is used for non-hydrogen atoms. As a consequence, jun-cc-pVTZ and aug-cc-pVDZ-F12 basis sets have comparable dimensions. Finally, it should be noted that for third-row atoms the variants including additional tight $d$-functions were employed for the Dunning-like basis sets, e.g. cc-pV($n$+$d$)Z or jun-cc-pV($n$+$d$)Z.\cite{1.1367373}

Aiming at setting up an approach affordable also for quite large systems, the reference geometry should be evaluated at an effective level of theory. As done in the definition of the junChS model,\cite{alessandrini2020} we rely on density functional theory (DFT) with dispersion corrections incorporated using Grimme’s D3 formulation. The double-hybrid revDSD-PBEP86-D3(BJ) functional\cite{revDSD,grimme2010consistent,grimme2011effect} in conjunction with the jun-cc-pVTZ basis set (hereafter referred to as revDSD) has been employed. Although in ref. \citenum{alessandrini2020} we have employed the B2PLYP functional,\cite{Grimme_B2PLYP_JCP06} the revDSD level is expected to provide improved results at the same computational cost.\cite{Barone20,Karton2021b} 

Focusing on structural determinations, the conventional junChS model (see ref. \citenum{alessandrini2020} for its definition) together with different ChS-F12 and full CCSD(T)-F12 composite schemes will be tested and their performance evaluated against the geometries reported by Hobza in ref. \citenum{Hobzagold}. 
For the geometry optimizations carried out with Molpro, the RMS of the displacements and gradients have been set to \num{4e-6} and \num{1e-6} a.u., respectively. These thresholds correspond to the ``very tight'' criteria used in the geometry optimizations performed with Gaussian.

\subsection{ChS Models: Dataset and Reference Energies}

\begin{figure}[b!]
    \centering
    \resizebox{15cm}{!}{\includegraphics{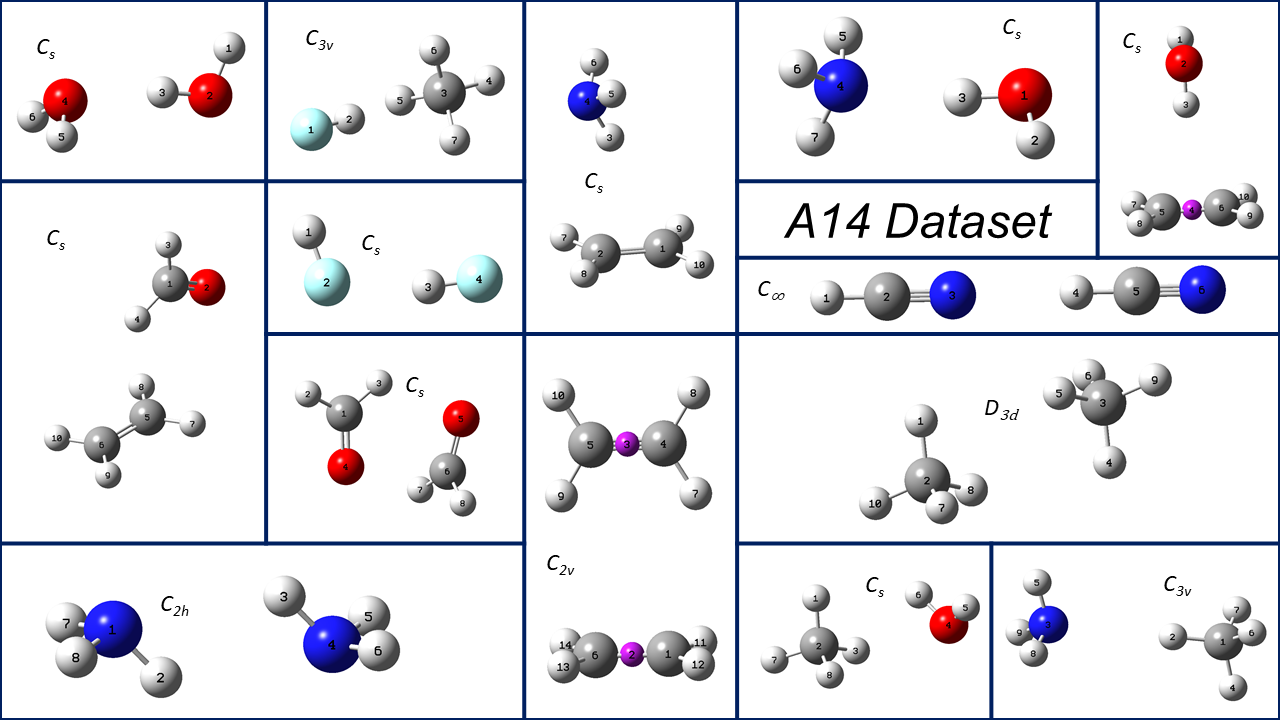}}
    \caption{The 14 small-sized non-covalent molecular complexes of the A24 dataset used in the present work. The symmetry point group is also provided.}
    \label{fig:figComplessi}
\end{figure}

The selection of the systems to be considered in this study has been based on the A24 dataset,\cite{Hobzagold} which is composed by 24 small non-covalent complexes covering strong interactions, such as hydrogen bonds, but also weaker contacts dominated by dispersion terms. Our interest mainly being the development of efficient strategies for the accurate description of large systems with a biological character, a subset of 14 complexes (hereafter referred to as A14 set) has been selected from the A24 dataset mentioned above. The systems containing either a noble gas, borane, ethyne, or ethane have been removed, as previously done in the definition of the junChS approach.\cite{alessandrini2020} The chosen systems are reported in Figure \ref{fig:figComplessi} and they are: 
\begin{itemize}
    \item Homo-dimers: \ce{H2O\bond{...}H2O}, \ce{NH3\bond{...}NH3}, \ce{CH2O\bond{...}CH2O}, \ce{HF\bond{...}HF}, \ce{HCN\bond{...}HCN}, \ce{CH4\bond{...}CH4}, \ce{C2H4\bond{...}C2H4};
    \item Hetero-dimers: \ce{CH4\bond{...}H2O}, \ce{NH3\bond{...}H2O}, \ce{CH4\bond{...}NH3}, \ce{CH4\bond{...}HF}, \ce{C2H4\bond{...}H2O}, \ce{C2H4\bond{...}NH3}, \ce{C2H4\bond{...}CH2O}.
\end{itemize}

For these systems, accurate interaction energies are available from the A24 dataset\cite{Hobzagold}. They have been obtained using a CCSD(T)-based composite scheme that accounts for the extrapolation to the CBS limit within the fc approximation together with the extrapolation to the CBS limit of the core-valence (CV) correlation energy.\cite{C5CP03151F} Therefore, this level of theory can be denoted as CCSD(T)/CBS+CV, or more simply CBS+CV. While more accurate values incorporating up to quadruple excitations in the cluster expansion and relativistic effects are available,\cite{C5CP03151F} one aim of this work is to provide an effective strategy to recover the error associated to the truncation of the basis set while accounting for the contribution of all electrons on top of fc-CCSD(T) calculations.\cite{Kesharwani2018b} Therefore, the CBS+CV level will be also designed as ``reference'', or more shortly as ``ref'', in the following. Instead, we will denote the CCSD(T)/CBS level,\cite{Hobzagold} i.e. the scheme not accounting for CV corrections, as ``ref-CBS''. 

\begin{figure}[b!]
\centering
\begin{subfigure}[b]{0.496\textwidth}
   \includegraphics[width=\textwidth]{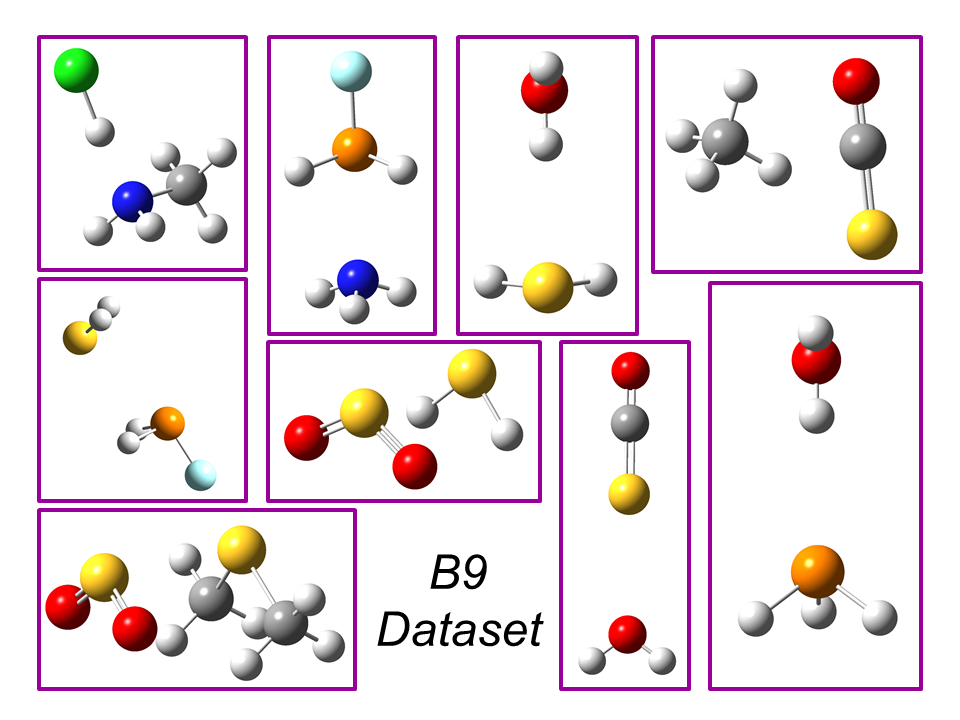}
   \caption{}
   \label{SecondoP}
\end{subfigure}
\begin{subfigure}[b]{0.496\textwidth}
   \includegraphics[width=\textwidth]{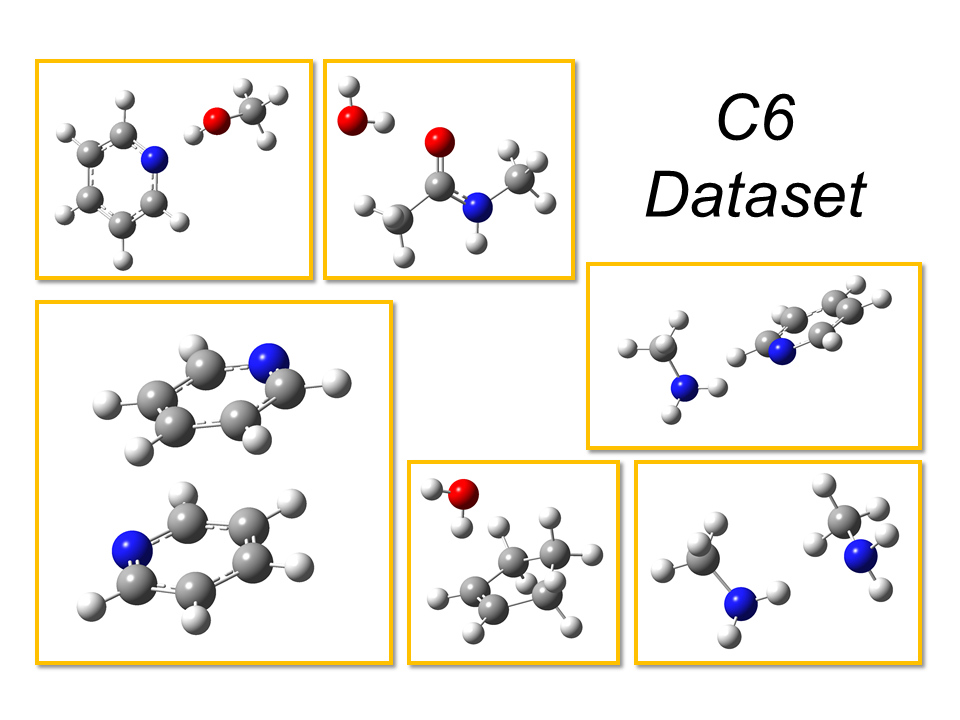}
   \caption{}
   \label{PrimoP} 
\end{subfigure}
\caption[Complexes]{Additional non-covalent complexes considered. Panel (a), B9 dataset: small systems containing atoms of the first three rows of the periodic table. Panel (b), C6 dataset: larger systems containing atoms of the first two rows of the periodic table}.
\end{figure}

While our conventional and explicitly correlated ChS schemes use revDSD geometries, in order to investigate possible structural effects, the reference geometries employed in the ``ref'' CBS+CV calculations presented above will be also exploited. We recall that they were obtained by means of a composite scheme that starts from the HF-SCF/aug-cc-pVQZ level of theory and incorporates the extrapolation to the CBS limit at the MP2 level using aug-cc-pV$n$Z basis sets ($n$=T,Q) and the contribution of triple excitations via CCSD(T) in combination with the aug-cc-pVDZ basis set. To rule out the basis set superposition error (BSSE) effects, the counterpoise (CP) correction\cite{boys1970calculation} was included at each iteration step. The resulting geometry will be denoted as ``CBS-georef'' in the following.

Since the A14 (but also A24) data set only includes species with atoms of the first two rows of the periodic table, a series of small complexes bearing at least one third-row atom (referred to as B9 set) has been added to the set of molecular complexes employed in this work. Furthermore, to test the new approach for larger non-covalent complexes, we have also built the C6 dataset. While a summary of these systems is provided here below, their graphical representation is reported in Figures~\ref{SecondoP} and~\ref{PrimoP}, respectively: 
\begin{itemize}
    \item B9 dataset: \ce{FH2P\bond{...}NH3}, \ce{FH2P\bond{...}H2S}, \ce{H2O\bond{...}H2S}, \ce{H2O\bond{...}PH3}, \ce{OCS\bond{...}H2O}, \ce{SO2\bond{...}H2S}, \ce{OCS\bond{...}CH4},  \ce{CH3NH2\bond{...}HCl}, \ce{(CH3)2S\bond{...}SO2}. 
    \item C6 dataset: \ce{c-C5H8\bond{...}H2O} (Z isomer), \ce{CH3NH2\bond{...}C5H5N}, \ce{CH3OH\bond{...}C5H5N}, \ce{H2O\bond{...}C3H7NO}, \ce{C5H5N\bond{...}C5H5N}, \ce{CH3NH2\bond{...}CH3NH2}. 
\end{itemize}

A last comment is deserved about the sought accuracy for computed interaction energies. First, it should be taken into account that they span a quite large range: from the relatively strong hydrogen bonds (10-30 \SI{}{\kilo\joule\per\mol}) to the much weaker dispersion interactions (1-10 \SI{}{\kilo\joule\per\mol}). As a consequence, the so-called gold-standard absolute error of 0.2 \SI{}{\kilo\joule\per\mol} suggested in ref. \citenum{chemrev16} is fully satisfactory for the first class of interactions, but it is questionable for the weak ones. Therefore, we prefer to retain the target accuracy introduced in ref. \citenum{alessandrini2020}: an average relative error within 1$\%$ without any outlier above 3$\%$, and an average absolute deviation below 0.2 \SI{}{\kilo\joule\per\mol}. 

\subsection{ChS Models: The Glossary}

To guide the reader through the plethora of ChS and ChS-F12 models that will be addressed in this manuscript, Table~\ref{glossary} summarizes and explains the corresponding acronyms. These can be seen as formed by three different pieces. The central part refers to the type of composite schemes. In the final part, the indication whether F12 methods are used is given, while in the initial part information on the basis set is provided. Whenever the type of basis set is not specified, it is implied the use of basis sets specifically developed for F12 computations (e.g. cc-pV$n$Z-F12) in combination with F12 methods and the use of the standard Dunning basis sets (e.g. cc-pV$n$Z) together with conventional methods. If not otherwise stated, the extrapolation to the CBS limit is performed using triple- and quadruple-zeta basis sets when conventional basis sets are employed, double-zeta and triple-zeta whenever F12 sets are used. In combination with the latter, the CV term is evaluated using the cc-pCVDZ-F12 basis set,\cite{1.3308483} while for schemes employing conventional basis sets, the CV contribution is computed with the cc-pwCVTZ basis set.\cite{peterson2002} Finally, the +jun$\Delta\alpha$ and +aug$\Delta\alpha$ labels indicate the additive inclusion of the contribution due to diffuse functions. In the corresponding models, the ChS scheme outlined above is augmented by a term evaluated as the difference between two MP2 computations, one using the aug-cc-pVTZ (or jun-cc-pVTZ) basis set and the other performed with cc-pVTZ.

\begin{table}[t!]
\centering
\caption{Glossary of conventional and F12 composite schemes: acronyms and their explanation.}
\label{glossary}
\resizebox{\textwidth}{!}{
\begin{tabular}{llcll}
\toprule
\multicolumn{2}{c}{Conventional methods} && \multicolumn{2}{c}{F12 methods} \\
\cline{1-2} \cline{4-5}
Acronym & \multicolumn{1}{c}{Explanation} && Acronym & \multicolumn{1}{c}{Explanation} \\ \midrule
\textbf{CBS} & CCSD(T) extrapolated && \textbf{CBS-F12} & CCSD(T)-F12 extrapolated \\
    & to the CBS limit     &&         & to the CBS limit\\
    & (cc-pV$n$Z)            &&         & (cc-pV$n$Z-F12) \\
\textbf{junCBS} & CCSD(T) extrapolated && \textbf{junCBS-F12} & CCSD(T)-F12 extrapolated \\
    & to the CBS limit using  &&         & to the CBS limit using \\
    & jun-cc-pV$n$Z basis sets  &&         & jun-cc-pV$n$Z basis sets \\
\textbf{CBS+CV} & CCSD(T) extrapolated && \textbf{CBS+CV-F12} & CCSD(T)-F12 extrapolated \\
    & to the CBS limit and    &&         & to the CBS limit and \\
    & CCSD(T)/CV correction   &&         & CCSD(T)-F12/CV correction   \\
    & (cc-pwCVTZ basis)       &&         & (cc-pCV$n$Z-F12 basis)          \\
\textbf{junCBS+CV} & CCSD(T) extrapolated && \textbf{junCBS+CV-F12} & CCSD(T)-F12 extrapolated \\
    & to the CBS limit using  &&         & to the CBS limit using \\
    & jun-cc-pV$n$Z basis sets  &&         & jun-cc-pV$n$Z basis sets \\
    & and CCSD(T)/CV corr.    &&         & and CCSD(T)-F12/CV corr.   \\
    & (cc-pwCVTZ basis)       &&         & (cc-pwCVTZ basis)          \\
\textbf{ChS} & ``cheap'' scheme with   && \textbf{ChS-F12} & ``cheap'' scheme using   \\
    & conventional methods    &&         & CCSD(T)-F12 and MP2-F12 \\
    & (cc-pV$n$Z / cc-pwCVTZ)   &&         & (cc-pV$n$Z-F12 / cc-pCV$n$Z-F12)   \\
\textbf{mayChS} & ChS scheme using     && \textbf{mayChS-F12} & ChS-F12 scheme using      \\
    & may-cc-pV$n$Z basis sets  &&         & may-cc-pV$n$Z basis sets \\
\textbf{junChS} & ChS scheme using     && \textbf{junChS-F12} & ChS-F12 scheme using      \\
    & jun-cc-pV$n$Z basis sets  &&         & jun-cc-pV$n$Z basis sets \\
\textbf{augChS} & ChS scheme using     && \textbf{augChS-F12} & ChS-F12 scheme using      \\
    & aug-cc-pV$n$Z basis sets  &&         & aug-cc-pV$n$Z-F12 basis sets \\
\textbf{augF12ChS} & ChS scheme using  \\
    & aug-cc-pV$n$Z-F12 basis sets  \\
\textbf{ChS+aug$\Delta\alpha$} & ChS scheme plus \\
    & additive contribution for \\
    & diffuse function using \\
    & aug-cc-pVTZ \\
\textbf{ChS+jun$\Delta\alpha$} & ChS scheme plus \\
    & additive contribution for \\
    & diffuse function using \\
    & jun-cc-pVTZ \\
\bottomrule
\end{tabular}
}
\end{table}

\subsection{ChS Models: Explicitly-Correlated F12 Variants}

While the conventional ChS approaches have already been introduced and applied,\cite{puzzarini2011extending,puzzarini2014accurate,li2018theory,alessandrini2020,puzzarini_jpc2020,alonso21} the explicitly correlated ChS models are here defined for the first time. As already noted for the conventional ChS approaches, different variants can be defined for the ChS-F12 model, with the general expression given by the following:
\begin{equation}
E_{ChS-F12}=E_{\mathrm{CC-F12}} + \Delta E_{CBS}\mathrm{(MP2-F12)} + \Delta E_\mathrm{CV}\mathrm{(MP2-F12)}	\; .
\label{cheap-f12}
\end{equation}

The starting fc-CCSD(T)-F12 energy, $E_{\mathrm{CC-F12}}$, is computed in conjunction with different basis sets, which are also employed in the CBS term. According to the basis sets used in the CCSD(T)-F12 and MP2-F12/CBS calculations, different schemes are obtained as summarized in Table~\ref{glossary}.
Let us see in details the expressions for the ChS-F12 and junChS-F12 models.
For the former, the CCSD(T)-F12 term is computed in combination with the cc-pVDZ-F12 basis set, while in the latter it is evaluated using the jun-cc-pVTZ set. 

The extrapolation to the CBS limit is performed using the $n^{-3}$ extrapolation formula:\cite{Helgaker97}  
\begin{equation}
    \Delta E_{CBS}\mathrm{(MP2-F12)} = \frac{n^{3}E(\mathrm{MP2-F12}/n\mathrm{Z})-(n-1)^{3}E(\mathrm{MP2-F12}/(n-1)\mathrm{Z})}{n^{3}-(n-1)^{3}} \; ,
    \label{cbs-cheap-f12}
\end{equation}
where ``$n$Z'' and ``($n$-1)Z'' stand, for example, for jun-cc-pVQZ and jun-cc-pVTZ, respectively, in the case of the junChS-F12 model, and for cc-pVTZ-F12 and cc-pVDZ-F12, respectively, when considering the ChS-F12 scheme. Noted is that ``($n$-1)Z'' is the basis set used in the corresponding fc-CCSD(T)-F12 calculation.

The CV contribution is then computed as difference between all-electron (ae) and fc MP2-F12 energy calculations:
\begin{equation}
    \Delta E_{CV}\mathrm{(MP2-F12)} = E(\mathrm{aeMP2-F12}/\mathrm{C}n\mathrm{Z}) - E(\mathrm{fcMP2-F12}/\mathrm{C}n\mathrm{Z}) \; ,
    \label{cv-cheap-f12}
\end{equation}
where ``C$n$Z'' stands for cc-pwCVTZ for all schemes not employing F12-basis sets (e.g. junChS-F12), while for the latter the cc-pCVDZ-F12 basis set is used.

A special note is deserved for the ChS-F12 model. In fact, while the standard model employs F12 double- and triple-zeta basis sets, the variant using F12 triple- and quadruple-zeta sets is also defined. In the latter case, the CCSD(T)-F12 term is computed with the cc-pVTZ-F12 set, the MP2-F12 extrapolation to the CBS limit using cc-pVTZ-F12 and cc-pVQZ-F12, and the MP2-F12 CV contribution with cc-pCVTZ-F12. 

In addition to the different ChS-F12 models generically defined above and summarized in Table~\ref{glossary}, composite schemes entirely based on CCSD(T)-F12 calculations and accounting for the extrapolation to the CBS limit and CV correction have also been defined and tested. According to Table~\ref{glossary}, these are defined as CBS+CV-F12. However, we need to specify the basis sets employed. If the cc-pVDZ-F12 and cc-pVTZ-F12 sets are used for the extrapolation and cc-pCVDZ-F12 for the CV term, then we obtain the CBS2+CV-F12 scheme. Otherwise, if the extrapolation to the CBS limit is performed using cc-pVTZ-F12 and cc-pVQZ-F12 and cc-pCVTZ-F12 is employed for the CV correction, the scheme is denoted as CBS3+CV-F12. If not explicitly specified, CBS+CV-F12 refers to CBS2+CV-F12. Finally, if the may-cc-pV$n$Z and jun-cc-pV$n$Z sets are used, we refer to it as mayCBS+CV-F12 and junCBS+CV-F12, respectively. In these latter cases, the CV contribution is evaluated using the cc-pwCVTZ basis set. If the CV term is neglected, the model is defined as CBS (conventional) and CBS-F12 (either CBS2-F12 or CBS3-F12; if not specified, we refer to the former).

\subsection{ChS Models: Structural Determination}

While the conventional ChS schemes have already been employed in structural determinations, their application to geometry optimizations of molecular complexes has not been introduced yet. As mentioned in the Introduction, this is accomplished in the present work for both conventional and explicitly correlated ChS schemes. The resulting composite approaches are denoted as ``geometry'' schemes and require geometry optimizations at different levels of theory. All the defined schemes do not account for the CP correction in view of its very small contribution, especially for explicitly-correlated approaches, as will be demonstrated in the results and discussion section.

The additivity of the various terms is exploited at a parameter level. For a generic structural parameter $R$, its expression is given by:
\begin{equation}
\label{chsR}
    R\mathrm{(ChS[-F12])} = R\mathrm{(CCSD(T)[-F12])} + \Delta R (\mathrm{MP2[-F12]/CBS}) + \Delta R(\mathrm{MP2[-F12]/CV}) \; ,
\end{equation}
where the first term on the right-hand side is the structural parameter $R$ optimized using CCSD(T) (or CCSD(T)-F12 for explicitly-correlated schemes) in conjunction with the specific basis set that defines the type of approach. To give a couple of examples, for the junChS model, this first term is the result of the fc-CCSD(T)/jun-cc-pVTZ geometry optimization, while it is the fc-CCSD(T)-F12/cc-pVDZ-F12 optimized parameter in the case of ChS-F12. The second term is the contribution due to the extrapolation to the CBS limit, which is derived from the extrapolation of the geometrical parameters obtained from fc-MP2 or MP2-F12 optimizations with two members of the hierarchical series of bases defining the scheme. This means jun-cc-pVTZ and jun-cc-pVQZ for the junChS model and cc-pVDZ-F12 and cc-pVTZ-F12 for the ChS-F12 approach. As done for energetics, the extrapolation is performed using the $n^{-3}$ expression as follows: 
\begin{equation}
\label{Rcbs}
    \Delta R (\mathrm{CBS}) = \frac{n^{3}R_{\mathrm{MP2[-F12]}/n\mathrm{Z}}-(n-1)^{3}R_{\mathrm{MP2[-F12]}/(n-1)\mathrm{Z}}}{n^{3}-(n-1)^{3}} - R_{\mathrm{MP2[-F12]}/(n-1)\mathrm{Z}} \; ,
\end{equation}
where $n$ = 4 (i.e. QZ), with the only exception of the ChS-F12 variant.  
The last term in equation \ref{chsR} introduces the effects on the geometry arising from the correlation of core electrons. This contribution is usually non negligible because equilibrium structures are particularly sensitive to the correlation of the inner electrons. The $\Delta R$({CV}) term is given by the difference between the structural parameter $R$ resulting from a geometry optimization with all electrons correlated and that within the frozen-core approximation. The basis sets used are: (i) the cc-pwCVTZ basis set for all conventional models and for the F12 approaches not employing F12 basis sets; (ii) the cc-pCV$n$Z-F12 basis sets, with $n$ = D or T, for those methods exploiting the F12 family of basis sets. 

Starting from the ChS model, two additional schemes have been defined by incorporating the effect of diffuse functions via an additive $\Delta\alpha$ term, $\Delta R$(MP2/diff), which can be computed using the jun-cc-pVTZ or aug-cc-pVTZ basis set. Such an addition leads to the definition of the ChS+jun$\Delta\alpha$ and ChS+aug$\Delta\alpha$, respectively. The $\Delta R$(MP2/diff) term is evaluated as a difference in the structural parameters optimized within the fc approximation. For the jun$\Delta\alpha$ correction, we have:
\begin{equation}
\label{Dalpha}
\Delta R(\mathrm{MP2/diff}) = R(\mathrm{fc-MP2/junTZ}) - R(\mathrm{fc-MP2/TZ})  \; .
\end{equation}

The CBS+CV scheme, entirely based on CC calculations, has also been used with the additivity being again applied at the geometry parameter level. In addition to the CBS2+CV-F12 and CBS3+CV-F12 models, entirely similar in the definition to the corresponding energy schemes, the junCBS+CV-F12 and junCBS+CV models have been introduced: 
\begin{equation}
\label{junCBSCV} 
R(\mathrm{junCBS+CV[-F12])} = R\mathrm{(CCSD(T)[-F12]/CBS)} + \Delta R({CCSD(T)[-F12]/CTZ}) \;  ,
\end{equation}
where, in both schemes, the jun-cc-pVTZ and jun-cc-pVQZ sets are used for the extrapolation to the CBS limit and cc-pwCVTZ for the CV term.

\section{Results and Discussion}

In the following section, the performance of the various ChS-F12 models, based on the A14 dataset, for both interaction energy evaluations and structural determinations will be presented with the aim of defining the most effective scheme. As a byproduct of this investigation, a database of accurate structures and energies for the A14+B9+C6 molecular complexes will be defined. In passing, the performance of the revDSD level of theory for both geometries and energies will be discussed.

\subsection{Performance of the ChS-F12 Models: Interaction Energy}

Together with the various ChS-F12 models (see Table~\ref{glossary}), the conventional junChS scheme has also been considered for comparison purposes and the role of the different contributions has been investigated. As mentioned in the the Methodology section, while the revDSD structures are the reference geometries for all our schemes, the geometry effect has been studied by resorting to the ``CBS-georef'' reference geometries. Noted is that, in addition to CP-corrected interaction energies, half-corrected (half-CP) and non-corrected (NCP) CP results will be also discussed.

\begin{table}[t!]
\centering
\caption{Relative and absolute errors\textsuperscript{a} of the revDSD level, the conventional junChS and augF12ChS models, and all ChS-F12 variants.\textsuperscript{b} Absolute errors in \SI{}{\kilo\joule\per\mol}.}
\label{tab:ChSF12withCV}
\begin{tabular}{@{}lccc@{}}
\toprule
Model &  & Relative error & Absolute error \\ \midrule
\multirow{3}{*}{revDSD\textsuperscript{b}} & CP      & 6.34\% & 0.49 \\ 
                                           & NCP     & 3.59\% & 0.34 \\
                                           & half-CP & 3.11\% & 0.24 \\ \midrule
\multirow{3}{*}{junChS}                    & CP      & 1.38\% & 0.12  \\
                                           & NCP     & 2.06\% & 0.22  \\
                                           & half-CP & 1.68\% & 1.68  \\ \midrule
\multirow{3}{*}{augF12ChS}                 & CP      & 1.12\% & 0.14  \\
                                           & NCP     & 9.70\% & 0.79  \\
                                           & half-CP & 4.34\% & 0.33  \\ \midrule
\multirow{3}{*}{mayChS-F12}                & CP      & 1.23\% & 0.10
  \\
                                           & NCP     & 2.81\% & 0.19
  \\
                                           & half-CP & 1.93\% & 0.14
  \\ \midrule
\multirow{3}{*}{junChS-F12}                & CP      & 0.68\% & 0.05
  \\
                                           & NCP     & 1.10\% & 0.08
  \\
                                           & half-CP & 0.88\% & 0.06
  \\ \midrule
\multirow{3}{*}{jun-($d,f$)H\_ChS-F12}     & CP      & 0.65\% & 0.05
  \\
                                           & NCP     & 1.07\% & 0.08
  \\
                                           & half-CP & 0.84\% & 0.06
  \\ \midrule
\multirow{3}{*}{ChS-F12\textsuperscript{c}}& CP      & 2.17\% (0.93\%)  & 0.18 (0.09)
                         \\
                                           & NCP     & 1.20\% (0.86\%)  & 0.12 (0.09)
                         \\
                                           & half-CP & 1.07\% (0.82\%)  & 0.08 (0.09)
                         \\ \midrule
\multirow{3}{*}{augChS-F12}                & CP      & 0.67\% & 0.06
   \\
                                           & NCP     & 2.12\% & 0.21
   \\
                                           & half-CP & 1.01\% & 0.10
   \\    \bottomrule                                
\end{tabular}

\smallskip
\textsuperscript{a} Errors evaluated with respect to CBS+CV ``ref'' reference energies. See text. \\
\textsuperscript{b} For all schemes: revDSD reference geometries. \\
\textsuperscript{c} In parentheses, the results for the scheme employing the F12 triple- and quadruple-zeta basis sets. See text. \\
\end{table}

\begin{figure}[t!]
\centering
   \includegraphics[width=1\linewidth]{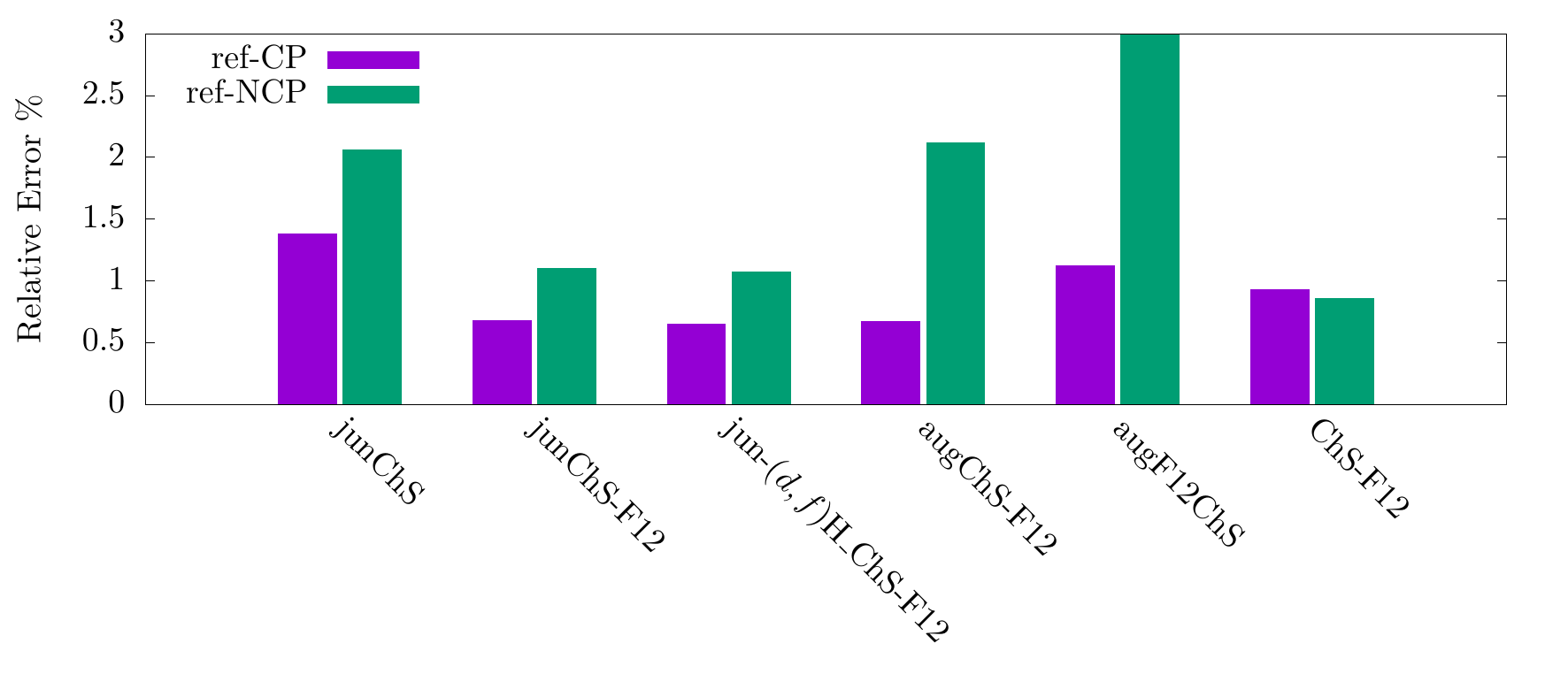}
\caption[Energy schemes]{CP and NCP energy differences with respect to reference, ``ref'', using revDSD geometries.}
\label{fig:istogramma}
\end{figure}

In Table \ref{tab:ChSF12withCV}, the results for the A14 dataset are summarized in terms of absolute and relative energies with respect to the ``ref'' values. The most promising F12 results are also graphically presented in Figure \ref{fig:istogramma}, where their relative errors are compared with those of the junChS counterpart. All F12 models of Table~\ref{glossary} are considered together with the conventional junChS and augF12ChS schemes, with the latter --despite being a conventional model-- employing the aug-cc-pVDZ-F12 and aug-cc-pVTZ-F12 basis sets. 
It is noted that the revDSD model provides large relative errors, these ranging from $\sim$3\% to $\sim$6\%. However, in absolute terms, the deviations are smaller than 0.5 \SI{}{\kilo\joule\per\mol}. The smallest mean absolute error (MAE) is delivered by the junChS-F12 model. Interestingly, reducing the dimension of the basis set by removing $d$ and $f$ polarization functions on hydrogen atoms (for triple- and quadruple-zeta basis sets, respectively), and thus resorting to the jun-($d,f$)H\_ChS-F12 model, even slightly improves the MAE. As far as the comparison with conventional schemes is concerned, it is evident that moving from the conventional junChS approach to the junChS-F12 counterpart nearly halves the MAE. However, the absolute errors point out that the differences between the two models, i.e. only fractions of \SI{}{\kilo\joule\per\mol}, are much less relevant with respect to what the MAEs tend to suggest. This is because half of the molecular complexes of the A14 set are characterized by small interaction energies. 

Back to Table~\ref{tab:ChSF12withCV}, the standard ChS-F12 scheme, which employs the F12 double- and triple-zeta basis sets, gives disappointing results and the same applies to the mayChS-F12 model. Indeed, $d$ and $f$ diffuse functions on non-hydrogen atoms (not included at these levels) play a non-negligible role. While the extension of the ChS-F12 scheme to the cc-pVTZ-F12 and cc-pVQZ-F12 sets (see numbers in parentheses and Figure \ref{fig:istogramma}) leads to improved results, this comes at the expense of a huge increase of the computational cost. Instead, the augChS-F12 model represents a much more effective alternative. Indeed, the jun-cc-pVTZ and aug-cc-pVDZ-F12 basis sets having comparable dimensions, it is somewhat natural that they lead to similar results. 

\begin{table}[t!]
\centering
\caption{Relative and absolute errors\textsuperscript{a} of different CBS+CV composite schemes.\textsuperscript{b} Absolute errors in \SI{}{\kilo\joule\per\mol}.}
\label{tab3}
\begin{tabular}{@{}lccc@{}}
\toprule
Model &  & Relative error & Absolute error \\ \midrule
\multirow{3}{*}{mayCBS+CV-F12}           &CP      & 0.87\%  & 0.08      \\
                                         &NCP     & 1.18\%  & 0.09      \\
                                         &half-CP & 0.92\%  & 0.08      \\ \midrule
\multirow{3}{*}{junCBS+CV-F12}           &CP      & 0.79\%  & 0.07      \\
                                         &NCP     & 0.92\%  & 0.09      \\
                                         &half-CP & 0.75\%  & 0.07      \\ \midrule
\multirow{3}{*}{jun-$(d,f)$H\_CBS+CV-F12}&CP      & 0.79\%  & 0.07      \\
                                         &NCP     & 0.99\%  & 0.10      \\
                                         &half-CP & 0.79\%  & 0.08      \\ \midrule
\multirow{3}{*}{CBS+CV-F12\textsuperscript{c}} &CP      & 1.36\%   & 0.10      \\
                                               &NCP     & 1.46\%   & 0.16      \\
                                               &half-CP & 1.00\%   & 0.08      \\ \bottomrule
\end{tabular}\\
\textsuperscript{a} Errors evaluated with respect to CBS+CV ``ref'' reference energies. See text. \\
\textsuperscript{b} For all schemes: revDSD reference geometries. \\
\textsuperscript{c} The extrapolation to the CBS limit using the $n^{-5}$ formula, with $n=3,4$.
\end{table}

In Table~\ref{tab3}, the results of a statistical analysis for some CBS+CV schemes entirely based on the CCSD(T)-F12 method is reported. In addition to the CBS+CV-F12, mayCBS+CV-F12, and junCBS+CV-F12 models, the approach employing the jun-cc-pV$n$Z-$(d,f)$H basis sets has also been considered. From the comparison of Tables~\ref{tab:ChSF12withCV} and~\ref{tab3}, it is apparent that, if we consider CP results, the ChS models perform even better than the CBS+CV schemes, the only exception being the approaches involving the may-cc-pV$n$Z basis sets. The situation is completely different if we analyze the NCP values. In fact, with the exception of the approaches involving the F12 basis sets, the best performance is obtained using the CBS+CV schemes. This behavior can be explained by a better extrapolation to the CBS limit in the full CCSD(T)-F12 schemes. Indeed, on a pure theoretical basis, CP and NCP values should be identical at the CBS limit because the BSSE should vanish. Therefore, CP correcting the energies should worsen the results, what actually occurs in the case of CBS+CV schemes. Since the MP2-F12 extrapolation is less effective, the CP correction within a ChS approach is still required and leads to improved results. Furthermore, CP correction in ChS schemes is computationally more convenient than in a CBS+CV scheme, the overall conclusion being that the junChS-F12 and jun-$(d,f)$H\_ChS-F12 models are the best performing schemes among the F12 approaches considered in this work.

\begin{table}[t!]
\centering
\caption{Relative and absolute errors\textsuperscript{a} of the CCSD(T)-F12/CBS approach in combination with different basis sets on top of the revDSD geometries. Absolute errors in \SI{}{\kilo\joule\per\mol}.}
\label{tab5}
\begin{tabular}{@{}lccccc@{}}
\toprule
      && \multicolumn{2}{c}{$n^{-3}$}     & \multicolumn{2}{c}{$n^{-5}$}      \\ \cline{3-4}\cline{5-6}
Model && Relative error & Absolute error  & Relative error & Absolute error   \\ \midrule
\multirow{3}{*}{cc-pV(D,T)Z-F12} & CP    & 0.86\% & 0.07  & 1.68\%  & 0.11    \\
                                 & NCP   & 1.77\% & 0.19  & 1.51\%  & 0.16    \\
                                 &half-CP& 1.17\% & 0.12  & 1.07\%  & 0.09    \\ \midrule
\multirow{3}{*}{cc-pV(T,Q)Z-F12} & CP    & 1.22\% & 0.13  & 0.77\%  & 0.08    \\
                                 & NCP   & 0.93\% & 0.11  & 0.85\%  & 0.11    \\
                                 &half-CP& 1.03\% & 0.12  & 0.81\%  & 0.09    \\ \midrule
\multirow{3}{*}{may-cc-pV(T,Q)Z} & CP    & 2.80\% & 0.24  & 0.81\%  & 0.08    \\
                                 & NCP   & 2.13\% & 0.14  & 1.23\%  & 0.10    \\
                                 &half-CP& 1.77\% & 0.15  & 0.93\%  & 0.08    \\ \midrule
\multirow{3}{*}{jun-cc-pV(T,Q)Z} & CP    & 1.64\% & 0.16  & 0.76\%  & 0.08    \\
                                 & NCP   & 1.21\% & 0.11  & 1.18\%  & 0.11    \\
                                 &half-CP& 1.25\% & 0.13  & 0.91\%  & 0.09    \\ \midrule
\multirow{3}{*}{jun-cc-pV(T,Q)Z-$d,f$H}& CP    & 1.80\%  & 0.17  & 0.77\%  &  0.08  \\
                                       & NCP   & 1.36\%  & 0.13  & 1.25\%  &  0.13  \\
                                       &half-CP& 1.44\%  & 0.15  & 0.95\%  &  0.10  \\ \bottomrule
\end{tabular}
\textsuperscript{a} Errors evaluated with respect to ``ref-CBS'' reference energies. See text.
\end{table}

\begin{table}[t!]
\centering
\caption{Relative and absolute errors\textsuperscript{a} of the extrapolation to the CBS limit within ChS approaches: $E_{\mathrm{CC-F12}} + \Delta E_{CBS}\mathrm{(MP2-F12)}$. Absolute errors in \SI{}{\kilo\joule\per\mol}.}
\label{tab6}
\begin{tabular}{@{}lccccc@{}}
\toprule
      && \multicolumn{2}{c}{$n^{-3}$}     & \multicolumn{2}{c}{$n^{-5}$}      \\ \cline{3-4}\cline{5-6}
Model && Relative error & Absolute error  & Relative error & Absolute error   \\ \midrule
\multirow{3}{*}{cc-pV(D,T)Z-F12}  & CP    & 2.27\%  & 0.18  & 2.88\%  & 0.23  \\
                                  & NCP   & 1.10\%  & 0.10  & 1.35\%  & 0.18  \\
                                  &half-CP& 1.30\%  & 0.09  & 1.62\%  & 0.15  \\ \midrule
\multirow{3}{*}{cc-pV(T,Q)Z-F12}  & CP    & 0.73\%  & 0.07  & 0.70\%  & 0.09  \\
                                  & NCP   & 0.90\%  & 0.09  & 0.79\%  & 0.06  \\
                                  &half-CP& 0.75\%  & 0.07  & 0.71\%  & 0.07  \\ \midrule
\multirow{3}{*}{may-cc-pV(T,Q)Z}  & CP    & 1.25\%  & 0.10  & 1.90\%  & 0.14  \\
                                  & NCP   & 2.87\%  & 0.19  & 1.59\%  & 0.09  \\
                                  &half-CP& 2.00\%  & 0.14  & 1.65\%  & 0.10  \\ \midrule
\multirow{3}{*}{jun-cc-pV(T,Q)Z}  & CP    & 0.66\%  & 0.05  & 1.03\%  & 0.07  \\
                                  & NCP   & 1.05\%  & 0.07  & 1.17\%  & 0.10  \\
                                  &half-CP& 0.84\%  & 0.06  & 0.77\%  & 0.05  \\ \midrule
\multirow{3}{*}{jun-cc-pV(T,Q)Z-$d,f$H}& CP    & 0.72\%  & 0.06  & 1.12\%  & 0.09  \\
                                       & NCP   & 1.02\%  & 0.07  & 1.09\%  & 0.08  \\
                                       &half-CP& 0.82\%  & 0.06  & 0.76\%  & 0.05  \\ \bottomrule
\end{tabular}
\textsuperscript{a} Errors evaluated with respect to ``ref-CBS'' reference energies. See text.
\end{table}

\subsubsection{Extrapolation to the CBS limit}

In view of the discussion above, it is interesting to investigate the convergence to the CBS limit, thereby resorting to the ``ref-CBS'' interaction energy as reference. In Tables~\ref{tab5} and~\ref{tab6}, the extrapolation to the CBS limit within CBS+CV-F12 and ChS-F12 schemes is analyzed, respectively, with the cc-pV$n$Z-F12, may-cc-pV$n$Z, jun-cc-pV$n$Z and jun-cc-pV$n$Z-$(d,f)$H basis sets being considered. Furthermore, the exponent $m$ of the $n^{-m}$ extrapolation has been tested. Among different preliminary investigations, we selected the $m$ = 5 exponent together with the standard $m$ = 3 exponent. From the CCSD(T)-F12/CBS results, summarized in terms of absolute and relative errors, it is noted that the MAEs of the NCP-mayCBS-F12 and NCP-junCBS-F12 interaction energies drop from 2.13\% and 1.21\% to 1.23\% and 1.18\%, respectively, when moving from $n^{-3}$ to $n^{-5}$. The reduction of the MAE is even more marked when considering the CP interaction energies; indeed, they reduce from 2.80\% and 1.64\% to 0.81\% and 0.76\%, respectively, going from $n^{-3}$ to $n^{-5}$. Analogous reductions are noted also for the jun-cc-pV$n$Z-$(d,f)$H and cc-pV$n$Z-F12 families of basis sets when considering the (T,Q) combination. For the (D,T) combination the CP-energies show instead the opposite behavior, possibly due to an error compensation related to the small basis set. 

From the inspection of Tables~\ref{tab5} and~\ref{tab6}, it appears that the $n^{-5}$ formula performs better than the $n^{-3}$ counterpart for the extrapolation of CCSD(T)-F12 energies. Instead, concerning MP2-F12 energies within the ChS-F12 models, we note that the conventional $n^{-3}$ extrapolation formula performs better than the $n^{-5}$ form, the only exception being the (T,Q) combination for the cc-pV$n$Z-F12 basis sets. Therefore, $n^{-3}$ has been retained for ChS-F12 models, while $n^{-5}$ has been chosen as the extrapolation of CCSD(T)-F12 energies.

\subsubsection{The Role of the Various ChS-F12 Contributions}

To discuss the different contributions within a given composite scheme, the best ChS-F12 variants have been selected, namely junChS-F12 and jun-($d,f$)H\_ChS-F12. For comparison purposes, the junCBS+CV-F12 model has also been considered. The results are collected in Table~\ref{1} for junChS-F12, Table~\ref{tab} for  jun-($d,f$)H\_ChS-F12, and Table~\ref{5} for junCBS+CV-F12. In all cases, CP-corrected energies are reported, the NCP counterparts being available in the SI (Tables \ref{tab:ncpjChSF12_si}, \ref{tab:ncpjChSF12df_si}, \ref{tab:ncpjCBSF12_si}). In all tables, CC stands for fc-CCSD(T)-F12 and, when present, MP2 stands for MP2-F12.  

\begin{table}[t!]
\centering
	\caption{junChS-F12 CP-energies (\SI{}{\kilo\joule\per\mol}): the various contributions for the A14 complexes.}
	\label{1}
	\resizebox{\textwidth}{!}{
\begin{tabular}{llcccccc} \toprule
    Complex  & ``ref''    & CC/junTZ & \multicolumn{1}{l}{$\Delta$MP2$^{\infty}$/jun(T,Q)Z} & MP2-CV/wCTZ  & \multicolumn{1}{l}{Total}    & \multicolumn{1}{l}{Rel. error (\%)} & \multicolumn{1}{l}{Abs. error} \\ \midrule
\ce{H2O\bond{...}H2O}       & -21.0832 & -20.8822                    & 0.0344                                            & -0.1512              & -20.9990                     & -0.40                              & 0.08                      \\
\ce{NH3\bond{...}NH3}        & -13.2131 & -12.9057                    & -0.2132                                           & -0.0807              & \multicolumn{1}{l}{-13.1997} & -0.10                              & 0.01                      \\
\ce{HF\bond{...}HF}                & -19.2213 & -19.1430                    & 0.0014                                            & -0.1078              & -19.2494                     & 0.15                               & -0.03                     \\
\ce{CH2O\bond{...}CH2O}      & -18.9284 & -18.5310                    & -0.3690                                           & -0.0552              & -18.9553                     & 0.14                               & -0.03                     \\
\ce{HCN\bond{...}HCN}             & -19.9828 & -19.7537                    & -0.0085                                           & -0.0787              & -19.8410                     & -0.71                              & 0.14                      \\
\ce{C2H4\bond{...}C2H4}& -4.6024  & -4.3114                     & -0.3000                                           & -0.0493              & -4.6607                      & 1.27                               & -0.06                     \\
\ce{CH4\bond{...}CH4}       & -2.2301  & -1.9832                     & -0.2083                                           & -0.0055              & -2.1970                      & -1.48                              & 0.03                      \\
\ce{H2O\bond{...}NH3}       & -27.3759 & -27.1443                    & -0.1294                                           & -0.2003              & -27.4740                     & 0.36                               & -0.10                     \\
\ce{H2O\bond{...}C2H4}    & -10.7696 & -10.4381                    & -0.2219                                           & -0.1092              & -10.7692                     & -0.004                             & 0.0004                    \\
\ce{C2H4\bond{...}CH2O}   & -6.7948  & -6.4974                     & -0.2677                                           & -0.0620              & -6.8271                      & 0.48                               & -0.03                     \\
\ce{NH3\bond{...}C2H4}    & -5.7865  & -5.5505                     & -0.2083                                           & -0.0603              & -5.8190                      & 0.56                               & -0.03                     \\
\ce{HF\bond{...}CH4}            & -6.9162  & -6.7403                     & -0.1943                                           & -0.1072              & -7.0418                      & 1.82                               & -0.13                     \\
\ce{H2O\bond{...}CH4}        & -2.8242  & -2.6431                     & -0.0979                                           & -0.0335              & -2.7746                      & -1.76                              & 0.05                      \\
\ce{NH3\bond{...}CH4}         & -3.2175  & -3.0678                     & -0.1160                                           & -0.0428              & -3.2266                      & 0.28                               & -0.01                     \\ \midrule
MAE       &          & \multicolumn{1}{l}{}        & \multicolumn{1}{l}{}                              & \multicolumn{1}{l}{} & \multicolumn{1}{l}{}         & 0.68                               & 0.05     \\          \bottomrule     
\end{tabular}
}
\end{table}

\begin{table}[ht]
\centering
\caption{jun-$(d,f)$H\_ChS-F12 CP-energies (\SI{}{\kilo\joule\per\mol}): the various contributions for the A14 complexes.
}
\label{tab}
\resizebox{\textwidth}{!}{%
\begin{tabular}{llcccccc}
\toprule
          & ``ref"      & CC/junTZ-$(d,f)$H & $\Delta$MP2$^{\infty}$/jun(T,Q)Z-$(d,f)$H & MP2-CV/wCTZ  & Total & Rel. error (\%) & Abs. error \\ \midrule
\ce{H2O\bond{...}H2O}   & -21.0832 & -20.8347                    & 0.0352                       & -0.1512 & -20.9506     & 0.63           & -0.13 \\
\ce{NH3\bond{...}NH3}   & -13.2131 & -12.8501                    & -0.2421                      & -0.0807 & -13.1729     & 0.30           & -0.04 \\
\ce{HF\bond{...}HF}     & -19.2213 & -19.0976                    & -0.0035                      & -0.1078 & -19.2089     & 0.06           & -0.01 \\
\ce{HCN\bond{...}HCN}   & -19.9828 & -19.7414                    & -0.0045                      & -0.0787 & -19.8246     & 0.79           & -0.16 \\
\ce{CH4\bond{...}CH4}   & -2.2301  & -1.9667                     & -0.2252                      & -0.0055 & -2.1974      & 1.47           & -0.03 \\
\ce{CH2O\bond{...}CH2O} & -18.9284 & -18.4796                    & -0.3971                      & -0.0552 & -18.9319     & 0.02           & 0.00  \\
\ce{C2H4\bond{...}C2H4} & -4.6024  & -4.2954                     & -0.3218                      & -0.0493 & -4.6665      & 1.39           & 0.06  \\
\ce{H2O\bond{...}C2H4}  & -10.7696 & -10.4093                    & -0.2320                      & -0.1091 & -10.7505     & 0.18           & -0.02 \\
\ce{H2O\bond{...}CH4}   & -2.8242  & -2.6338                     & -0.0999                      & -0.0339 & -2.7676      & 2.00           & -0.06 \\
\ce{H2O\bond{...}NH3}   & -27.3759 & -27.0710                    & -0.1545                      & -0.2002 & -27.4257     & 0.18           & 0.05  \\
\ce{NH3\bond{...}CH4}   & -3.2175  & -3.0571                     & -0.1161                      & -0.0423 & -3.2155      & 0.06           & 0.00  \\
\ce{NH3\bond{...}C2H4}  & -5.7865  & -5.5344                     & -0.2189                      & -0.0602 & -5.8136      & 0.47           & 0.03  \\
\ce{HF\bond{...}CH4}    & -6.9162  & -6.6717                     & -0.2239                      & -0.1072 & -7.0029      & 1.25           & 0.09  \\
\ce{C2H4\bond{...}CH2O} & -6.7948  & -6.4739                     & -0.2820                      & -0.0600 & -6.8159      & 0.31           & 0.02  \\ \midrule
MAE     &    &    &    &         &   & 0.65           & 0.05  \\ \bottomrule     
\end{tabular}%
}
\end{table}

\begin{table}[t!]
\centering
\caption{junCBS+CV-F12 CP-energies (\SI{}{\kilo\joule\per\mol}): the various contributions for the A14 complexes.} 
\label{5}
	\resizebox{\textwidth}{!}{
\begin{tabular}{llcccccc} \toprule
                               & ``ref"    & CC-CBS/jun(T,Q)Z & CC-CV/wCTZ  & Total  & \multicolumn{1}{l}{Rel. error (\%)} & \multicolumn{1}{l}{Abs. error} & \multicolumn{1}{l}{} \\ \midrule
\ce{H2O\bond{...}H2O}         & -21.0832 & -21.0524                    & -0.1200               & -21.1724 & 0.42                                    & -0.09                     &                      \\
\ce{NH3\bond{...}NH3}         & -13.2131 & -13.1601                    & -0.0605               & -13.2206 & 0.06                                    & -0.01                     &                      \\
\ce{HF\bond{...}HF}                & -19.2213 & -19.3738                    & -0.0818               & -19.4556 & 1.22                                    & -0.23                     &                      \\
\ce{CH2O\bond{...}CH2O}       & -18.9284 & -18.9566                    &  0.0382                & -18.9184  & -0.05                                   & 0.01                      &                      \\
\ce{HCN\bond{...}HCN}               & -19.9828 & -19.8441                    & -0.0811               & -19.9253 & -0.29                                   & 0.06                      &                      \\
\ce{C2H4\bond{...}C2H4} & -4.6024  & -4.5555                     & -0.0068               & -4.5623  & -0.87                                   & 0.04                      &                      \\
\ce{CH4\bond{...}CH4}        & -2.2301  & -2.2058                     &  0.0049                & -2.2009  & -1.31                                   & 0.03                      &                      \\
\ce{H2O\bond{...}NH3}         & -27.3759 & -27.4233                    & -0.1633               & -27.5865 & 0.77                                    & -0.21                     &                      \\
\ce{H2O\bond{...}C2H4}     & -10.7696 & -10.6645                    & -0.0532              & -10.7177 & -0.48                                   & 0.05                      &                      \\
\ce{C2H4\bond{...}CH2O}    & -6.7948  & -6.7604                     & -0.0010               & -6.7704  & -0.36                                   & 0.02                      &                      \\
\ce{NH3\bond{...}C2H4}     & -5.7865  & -5.7393                     & -0.0243               & -5.7636  & -0.40                                   & 0.02                      &                      \\
\ce{HF\bond{...}CH4}            & -6.9162  & -7.0176                     & -0.0749               & -7.0925  & 2.55                                    & -0.18                     &                      \\
\ce{H2O\bond{...}CH4}         & -2.8242  & -2.7537                     & -0.0206               & -2.7743  & -1.77                                   & 0.05                      &                      \\
\ce{NH3\bond{...}CH4}         & -3.2175  & -3.2065                     & -0.0274               & -3.2338  & 0.51                                    & -0.02                     &                      \\ \midrule
MAE      &          &                                            &                                     &              & 0.79                                    & 0.07                      \\              \bottomrule       
\end{tabular}
}
\end{table}

From Tables~\ref{1} and~\ref{tab}, we note that --in absolute terms-- the CCSD(T)-F12/jun-cc-pVTZ (or CCSD(T)-F12/jun-cc-pVTZ-$(d,f)$H) terms underestimate the ``ref'' values by a quantity ranging from $\sim$0.1 to $\sim$0.4 \SI{}{\kilo\joule\per\mol}. In almost all cases, about 70\% of these differences are recovered by the CBS correction term evaluated at the MP2-F12 level. This contribution is in almost all cases negative, which means that it increases --in absolute terms-- the interaction energies obtained at the CCSD(T)-F12 level in conjunction with the triple-zeta basis set. Furthermore, if the comparison is extended to Table~\ref{5}, we note that the CCSD(T)-F12 terms augmented by the MP2-F12/CBS corrections (Tables~\ref{1} and~\ref{tab}) are very close to the corresponding junCBS-F12 values. This means that the approximation of recovering the extrapolation to the CCSD(T)-F12 CBS limit at the MP2-F12 level is valid. 
Moving to the CV contribution, it is again noted that CCSD(T)-F12 and MP2-F12 provide similar corrections. In almost all cases, the CV corrections are negative, thus further increasing the computed interaction energies. Overall, the CV term is not negligible at all, indeed this being in almost all cases larger (up to one order of magnitude) than the absolute error associated to the model considered. 

Finally, it is interesting to observe that, if NCP-interaction energies are considered (see the SI), all the above observations remain valid, the only difference being that, for the junChS-F12 and jun-$(d,f)$H\_ChS-F12 schemes, the MAE increases from 0.68\% to 1.10\% and from 0.65\% to 1.07\%, respectively. Analogously, for the junCBS+CV-F12 model, the MAE increases from 0.79 \% to 0.92\%. 
However, despite these significant differences in the CP- and NCP-MAEs, in absolute terms, they are very small, indeed pointing out that --owing to the extrapolation to the CBS limit-- the CP corrections are small. If the comparison is extended to the conventional junChS model (see the SI, Tables \ref{tab:jchs_cp_si} and \ref{tab:jchs_ncp_si}), we note that the CP-NCP energy differences are slightly more pronounced for schemes involving conventional methods when comparing models employing the same basis sets (see Figure~\ref{fig:istogramma}). For example, the difference is larger for junChS than for junChS-F12. This is even more pronounced when comparing augF12ChS and augChS-F12 (both employing the aug-cc-pV$n$Z-F12 sets, see Table~\ref{glossary}). To conclude this subsection, it is worthwhile noting once again that the junChS-F12 and jun-$(d,f)$H\_ChS-F12 approaches considered in the detailed discussion above perform better than the junCBS+CV-F12 scheme despite their reduced computational cost.

\subsubsection{Geometry Effects on Interaction Energies}

\begin{table}[t!]
\centering
\caption{Relative and absolute errors\textsuperscript{a} of the revDSD level, the conventional junChS and the junChS-F12 variants employing ``CBS-georef'' reference geometries. Absolute errors in \SI{}{\kilo\joule\per\mol}.}
\label{tab:F12suREF}
\begin{tabular}{@{}lccc@{}}
\toprule
Model &  & Relative error & Absolute error \\ \midrule
\multirow{3}{*}{revDSD}               & CP    & 6.74\%   & 0.54  \\
                                      & NCP   & 3.31\%   & 0.29  \\
                                      &half-CP& 3.45\%   & 0.28  \\ \midrule
\multirow{3}{*}{junChS}               & CP    & 0.93\%   & 0.10  \\
                                      & NCP   & 1.35\%   & 0.14  \\
                                      &half-CP& 1.03\%   & 0.11  \\ \midrule
\multirow{3}{*}{junChS-F12}           & CP    & 0.67\%   & 0.073 \\
                                      & NCP   & 0.98\%   & 0.072 \\
                                      &half-CP& 0.82\%   & 0.071 \\ \midrule
\multirow{3}{*}{jun-($d,f$)H\_ChS-F12}& CP    & 0.74\%   & 0.083  \\
                                      & NCP   & 0.95\%   & 0.066  \\
                                      &half-CP& 0.83\%   & 0.073  \\ 
\bottomrule
\end{tabular}

\textsuperscript{a} Errors evaluated with respect to CBS+CV ``ref'' reference energies. See text. \\

\end{table}

In order to investigate the effect of the reference structures and possible errors associated to their choice, a comparison has been carried between NCP and CP interaction energies of the A14 complexes obtained on top of the ``CBS-georef'' and revDSD structures. In passing we note that all revDSD optimized structures belong to the same symmetry point group as those reported by Hobza,\cite{Hobzagold} with the revDSD structural parameters of A14 data set gathered in the SI (Table \ref{tab:geomcomplessi_si}). The results to be compared are those of Table~\ref{tab:ChSF12withCV} (revDSD geometry) and Table~\ref{tab:F12suREF} (CBS-georef geometry). 

We limit the discussion to the three most promising composite methods, namely junChS, junChS-F12 and junChS-F12-($d,f$)H. For them, when CP interaction energies are considered, the MAEs are 1.38\%, 0.68\%, and 0.65\% when using revDSD geometries, and 0.93\%, 0.67\%, and 0.74\% when employing the ``CBS-georef'' structures, respectively. Similar variations are noted for NCP interaction energies. We can thus conclude that improving the reference structure leaves essentially unchanged the performance of the jun-ChS-F12 scheme, while that of the conventional junChS is significantly improved.  A slight worsening is noted for the junChS-F12-($d,f$)H model. However, in absolute terms, these variations lead to negligible energy differences, well below 0.05 \SI{}{\kilo\joule\per\mol}.

\subsection{Performance of the ChS-F12 Models: Structural Determination}

The performance of the geometry schemes introduced in the methodology section has been tested using the molecular complexes of the A14 dataset. 
The most critical and sensitive parameter to analyze in the case of non-covalent complexes is the inter-molecular distance ruling the interaction and, indeed, this will be the quantity discussed in detail in the following. For the sake of completeness, all parameters describing the complexes are reported, for all the approaches considered, in the SI (Table \ref{tab:geomcomplessi_si}). The main comment on the intra-molecular distances is that they appear to be similar at the different levels of theory considered, with differences of few \SI{}{\milli\angstrom}. The same applies to the intra-molecular angles, which are quite insensitive to the level of theory and only changes smaller than 1 degree are observed. For example, in the dimer of water the two intra-molecular H-O-H angles are predicted to be 104.98 and 104.88 degrees using the junCBS+CV-F12 model. The same angles are 104.90 and 104.84 degrees, respectively, at the junChS level, and 104.72 and 104.92 degrees, respectively, from the junChS-F12 approach. The inter-molecular O4-H3-O2 angle spans from 171.15 to 172.68 degrees, the maximum variation between different composite schemes thus being 1.5 degrees. Since the PES along an inter-molecular angle is usually quite flat, one can conclude that such small changes lie within the uncertainty of the structural determination and that the overall energetics of the system is not influenced by small changes in intra-/inter-molecular angles. 
Another example to support the conclusion that intra-molecular distances and angles as well as inter-molecular angles are little affected by the model considered is offered by the \ce{CH4\bond{...}NH3} adduct. This is characterized by a H-C-H intra-molecular angle which is found to be 109.7 degrees by all the schemes employed. As before, the intra-molecular distances are very similar at all levels of theory, with differences well within \SI{4}{\milli\angstrom}. At the junCBS+CV-F12 level, the N-H and C-H bond lengths are 1.011 and 1.087 \AA, respectively. 
To monitor different types of complexes, we can also consider a system dominated by dispersion interactions, namely the methane dimer. In this case, the C-H intra-molecular distances obtained by exploiting different composite schemes are very similar (see SI for values) and the same applies to the inter-molecular angle, whose predicted value is always close to 70.4 degrees. 

Let us now focus the discussion on inter-molecular distances, which are instead particularly sensitive to the computational approach considered. The first question to address is the reliability of models employing an additive term for incorporating the effects of diffuse functions (the ChS+aug$\Delta \alpha$ or ChS+jun$\Delta \alpha$ schemes; see Table~\ref{glossary}). Let us consider a couple of examples. For the rather rigid \ce{NH3\bond{...}H2O} complex, the N$\cdots$O distance is \SI{1.957}{\angstrom} at the ChS+aug$\Delta \alpha$ level, to be compared with the value of about $\sim$ \SI{1.975}{\angstrom} delivered by the junChS and junCBS+CV-F12 models. The difference being close to 20 \SI{}{\milli\angstrom} leads to the conclusion that the ChS+aug$\Delta\alpha$ approach is not suitable for evaluating geometries of non-covalent complexes. The N$\cdots$O distance issuing from ChS+jun$\Delta\alpha$ geometry optimizations is 1.960 \SI{}{\angstrom}, thus reducing the difference with respect to junCBS+CV-F12 to 15 \SI{}{\milli\angstrom}, which is --however-- an unsatisfactory value. Moving to a more flexible system, i.e. the \ce{CH4\bond{...}NH3} complex, the inadequacy of the ChS+aug$\Delta \alpha$ and ChS+jun$\Delta \alpha$ schemes is even more evident: these models provide a N$\cdots$H distance about \SI{0.2}{\angstrom} longer than what predicted by the other composite approaches, irrespective of the considered family of basis sets. In conclusion, the additive approximation for the effect of diffuse functions is not suitable not only for interaction energies \cite{alessandrini2020}, but also for structural determinations. 

Having excluded the $\Delta\alpha$ composite schemes, we can now move to address the reliability of the other models. In general terms, the performances of the junChS and junChS-F12 variants are comparable, while the junCBS+CV-F12 approach gives typically different bond distances which can be either longer or shorter than the junChS and junChS-F12 counterparts. Analysing in detail the trends, one may observe that for some species the convergence to the CBS limit of the CCSD(T)-F12 method is not what expected. Indeed, for \ce{NH3\bond{...}H2O}, \ce{H2O\bond{...}C2H4}, \ce{HCN\bond{...}HCN}, and \ce{CH4\bond{...}HF}, the inter-molecular bond distance increases when going from the jun-cc-pVTZ to the jun-cc-pVQZ basis set, a behaviour which is opposite to that systematically observed for the conventional CCSD(T) and MP2 methods as well as for MP2-F12. From our analysis, it can be concluded that CCSD(T)-F12 geometries do not seem to benefit from extrapolation to the CBS limit, either with the $n^{-3}$ or $n^{-5}$ formula and, actually, the extrapolation can lead to unreliable corrections. The former point confirms the fact that F12 methodologies approach in a steep manner the CBS limit and that the convergence is achieved with relatively small basis sets. This is also confirmed by the fact that the MP2-F12 corrective term for incorporating the CBS limit in ChS-F12 schemes often leads to corrections smaller than 0.1 m\AA. In this respect, we can mention the \ce{CH4\bond{...}H2O} and \ce{C2H4\bond{...}NH3} complexes. Interestingly, the comparison between CCSD(T)-F12/jun-cc-pVTZ and junCBS+CV-F12 geometries points out that the difference is negligible or can be attributed to the CV term. This means that the simplest and ``safest" route to obtain accurate structural parameters when relying on CCSD(T)-F12 calculations is to add CV corrections, evaluated at either the MP2-F12 or CCSD(T)-F12 level, to CCSD(T)-F12/jun-cc-pVTZ geometries. 

From the inspection of Table~\ref{tab:geomcomplessi_si} of the SI, it is observed that the junChS and junChS-F12 models show the best performances, with the CBS-georef structures used as references. For inter-molecular distances, the average deviation from the latter is -0.01 \AA. Half of such a discrepancy can be attributed to the CV term, which is not included in the CBS-georef reference structures and, for the junChS and junChS-F12 approaches, indeed amounts --on average-- to -0.005 \AA. Even if not accounting for the diffuse functions effects, the ChS scheme shows a performance only little degraded with respect to junChS and junChS-F12. The CBS and CV contributions on the geometrical parameters within the ChS, junChS and junChS-F12 models are collected in Table~\ref{tab:cbsgeomparam_si} of the SI. From the inspection of this table, we note that --on average-- the CBS correction to the inter-molecular distances is larger for ChS and junChS than for junChS-F12. While for the two schemes involving conventional methods the CBS contribution is often on the order of a few hundredths of \AA, for junChS-F12 this term is generally smaller than 0.01 \AA. Furthermore, the CBS contribution is larger for inter-molecular distances than for the intra-molecular counterparts. Moving to the CV term, there is not a noticeable difference between intra- and inter-molecular parameters. In the case of distances, the correction is always on the order of a few m\AA. Interestingly, the revDSD geometries show a good accuracy, thus supporting the choice made in the definition of our ChS/ChS-F12 approaches. The average deviation, in absolute terms, from the CBS-georef structures is 0.016 \AA, with maximum discrepancies up to 0.03 \AA. While there is a systematic trend in the deviations of the junChS, junChS-F12, and ChS models from CBS-georef, the same does not apply to revDSD, the differences being either positive or negative.


\begin{table}[t!]
\centering
\caption{Comparison of CP and NCP corrected geometries for some paradigmatic cases.}
\label{tab:geometriecpncp}
\resizebox{!}{0.445\textheight}{
\begin{tabular}{@{}ccccccccccc@{}}
\toprule
                            &                        & \multicolumn{2}{c}{junChS} & \multicolumn{2}{c}{junChS-F12} & \multicolumn{2}{c}{junCBS+CV} & \multicolumn{2}{c}{junCBS+CV-F12\textsuperscript{a}} & CC-F12/VDZ-F12\textsuperscript{b} \\ \cline{3-10}
                            &                        & CP           & NCP         & CP             & NCP           & CP            & NCP           & CP               & NCP             & NCP                                                \\ \midrule
\multirow{20}{*}{\ce{H2O\bond{...}H2S}} & $r$(O1-H2)             & 0.9603       & 0.9603      & 0.9605         & 0.9606        & 0.9606        & 0.9612        & 0.9608           & 0.9608          & 0.9651                                             \\
                            &                        &              &             &                &               &               &               & (0.9603)         & (0.9603)        &                                                    \\
                            & $r$(O1-H3)             & 0.9560       & 0.9561      & 0.9563         & 0.9563        & 0.9563        & 0.9570        & 0.9566           & 0.9566          & 0.9608                                             \\
                            &                        &              &             &                &               &               &               & (0.9561)         & (0.9561)        &                                                    \\
                            & $\theta$(H2-O1-H3)     & 104.67       & 104.69      & 104.81         & 104.91        & 104.82        & 104.66        & 104.73           & 104.73          & 104.40                                             \\
                            &                        &              &             &                &               &               &               & (104.74)         & (104.75)        &                                                    \\
                            & $r$(H3$\cdots$S4)      & 3.4821       & 3.4758      & 3.4656         & 3.4716        & 3.4738        & 3.4799        & 3.4847           & 3.4836          & 3.4722                                             \\
                            &                        &              &             &                &               &               &               & (3.4805)         & (3.4850)        &                                                    \\
                            & $\theta$(O1-H3-S4)     & 116.22       & 116.69      & 119.32         & 117.78        & 116.61        & 116.48        & 116.27           & 116.33          & 116.83                                             \\
                            &                        &              &             &                &               &               &               & (116.27)         & (116.47)        &                                                    \\
                            & $r$(H5/H6-S4)          & 1.3363       & 1.3363      & 1.3367         & 1.3344        & 1.3369        & 1.3368        & 1.3366           & 1.3366          & 1.3391                                             \\
                            &                        &              &             &                &               &               &               & (1.3365)         & (1.3364)        &                                                    \\
                            & $\theta$(H5-S4-H3)     & 84.29        & 83.78       & 81.85          & 83.07         & 84.31         & 84.05         & 84.36            & 84.34           & 82.82                                              \\
                            &                        &              &             &                &               &               &               & (84.38)          & (84.21)         &                                                    \\
                            & $\varphi$(H5-S4-H3-O1) & 133.55       & 133.50      & 133.23         & 133.39        & 133.49        & 133.49        & 133.48           & 133.48          & 133.41                                             \\
                            &                        &              &             &                &               &               &               & (133.48)         & (133.47)        &                                                    \\
                            & $\varphi$(H6-S4-H3-O1) & -133.55      & -133.50     & -133.23        & -133.39       & -133.49       & -133.49       & -133.48          & -133.48         & -133.41                                            \\
                            &                        &              &             &                &               &               &               & (-133.48)        & (-133.47)       &                                                    \\
                            & $\theta$(H5-S4-H6)     & 92.46        & 92.29       & 92.42          & 92.33         & 92.43         & 92.38         & 92.45            & 92.44           & 92.23                                              \\
                            &                        &              &             &                &               &               &               & (92.46)          & (92.45)         &                                                    \\ \midrule
\multirow{20}{*}{\ce{H2O\bond{...}H2O}} & $r$(O2-H1)             & 0.9548       & 0.9554      & 0.9550         & 0.9556        & 0.9563        & 0.9563        & 0.9559           & 0.9559          & 0.9602                                             \\
                            &                        &              &             &                &               &               &               & (0.9554)         & (0.9554)        &                                                    \\
                            & $r$(O2-H3)             & 0.9625       & 0.9626      & 0.9631         & 0.9629        & 0.9634        & 0.9635        & 0.9631           & 0.9632          & 0.9672                                             \\
                            &                        &              &             &                &               &               &               & (0.9627)         & (0.9626)        &                                                    \\
                            & $\theta$(H3-O2-H1)     & 105.38       & 104.84      & 105.48         & 104.92        & 104.80        & 104.81        & 104.89           & 104.88          & 104.49                                             \\
                            &                        &              &             &                &               &               &               & (104.89)         & (104.89)        &                                                    \\
                            & $r$(H3$\cdots$O4)      & 2.1860       & 1.9507      & 2.1614         & 1.9487        & 1.9571        & 1.9513        & 1.9533           & 1.9513          & 1.9600                                             \\
                            &                        &              &             &                &               &               &               & (1.9494)         & (1.9512)        &                                                    \\
                            & $\theta$(O4-H3-O2)     & 168.35       & 171.15      & 169.18         & 171.24        & 171.78        & 171.41        & 171.79           & 171.82          & 171.73                                             \\
                            &                        &              &             &                &               &               &               & (171.93)         & (171.75)        &                                                    \\
                            & $r$(H5/H6-O4)          & 0.9554       & 0.9570      & 0.9567         & 0.9572        & 0.9579        & 0.9579        & 0.9575           & 0.9575          & 0.9618                                             \\
                            &                        &              &             &                &               &               &               & (0.9570)         & (0.9570)        &                                                    \\
                            & $\theta$(H5-O4-H3)     & 107.70       & 110.74      & 109.96         & 111.85        & 111.59        & 110.89        & 111.63           & 111.54          & 109.79                                             \\
                            &                        &              &             &                &               &               &               & (111.92)         & (111.66)        &                                                    \\
                            & $\varphi$(H5-O4-H3-O2) & 55.63        & 57.89       & 57.32          & 58.67         & 58.50         & 58.00         & 58.59            & 58.46           & 57.24                                              \\
                            &                        &              &             &                &               &               &               & (58.78)          & (58.54)         &                                                    \\
                            & $\varphi$(H6-O4-H3-O2) & -55.63       & -57.89      & -57.32         & -58.67        & -58.50        & -58.00        & -58.59           & -58.46          & -57.24                                             \\
                            &                        &              &             &                &               &               &               & (-58.78)         & (-58.54)        &                                                    \\
                            & $\theta$(H5-O4-H6)     & 104.25       & 104.90      & 103.98         & 104.72        & 104.89        & 104.88        & 104.98           & 104.98          & 104.61                                             \\
                            &                        &              &             &                &               &               &               & (105.01)         & (104.99)        &                                                    \\ \midrule
\multirow{8}{*}{\ce{CH4\bond{...}CH4}}  & $r$(C2-H1)             & 1.0857       & 1.0857      & 1.0857         & 1.0857        & 1.0864        & 1.0864        & 1.0860           & 1.0860          & 1.0883                                             \\
                            &                        &              &             &                &               &               &               & (1.0858)         & (1.0858)        &                                                    \\
                            & $r$(C2$\cdots$C3)      & 3.6233       & 3.6231      & 3.6397         & 3.6373        & 3.6435        & 3.6388        & 3.6403           & 3.6385          & 3.6526                                             \\
                            &                        &              &             &                &               &               &               & (3.6251)         & (3.6296)        &                                                    \\
                            & $\theta$(C3-C2-H1)     & 70.44        & 70.44       & 70.44          & 70.45         & 70.44         & 70.44         & 70.44            & 70.44           & 70.44                                              \\
                            &                        &              &             &                &               &               &               & (70.44)          & (70.44)         &                                                    \\
                            & $r$(C3-H9)             & 1.0857       & 1.0857      & 1.0857         & 1.0857        & 1.0864        & 1.0864        & 1.0860           & 1.0860          & 1.0883                                             \\
                            &                        &              &             &                &               &               &               & (1.0858)         & (1.0858)        &                                                    \\ \midrule
\multirow{10}{*}{\ce{HCN\bond{...}HCN}} & $r$(H1-C2)             & 1.0658       & 1.0656      & 1.0659         & 1.0658        & 1.0662        & 1.0661        & 1.0660           & 1.0659          & 1.0666                                             \\
                            &                        &              &             &                &               &               &               & (1.0660)         & (1.0658)        &                                                    \\
                            & $r$(C2-N3)             & 1.1505       & 1.1505      & 1.1500         & 1.1501        & 1.1518        & 1.1519        & 1.1507           & 1.1508          & 1.1560                                             \\
                            &                        &              &             &                &               &               &               & (1.1500)         & (1.1501)        &                                                    \\
                            & $r$(N3$\cdots$H4)      & 2.2159       & 2.2164      & 2.2160         & 2.2162        & 2.2192        & 2.2157        & 2.2156           & 2.2146          & 2.2204                                             \\
                            &                        &              &             &                &               &               &               & (2.2129)         & (2.2153)        &                                                    \\
                            & $r$(H4-C5)             & 1.0715       & 1.0714      & 1.0717         & 1.0716        & 1.0719        & 1.0719        & 1.0717           & 1.0717          & 1.0720                                             \\
                            &                        &              &             &                &               &               &               & (1.0717)         & (1.0717)        &                                                    \\
                            & $r$(C5-N6)             & 1.1526       & 1.1528      & 1.1521         & 1.1522        & 1.1540        & 1.1540        & 1.1529           & 1.1529          & 1.1581                                             \\
                            &                        &              &             &                &               &               &               & (1.1522)         & (1.1523)        &                                                    \\ \bottomrule
\end{tabular}
}

{\footnotesize
\textsuperscript{a} Extrapolations using the $n^{-3}$ formula are reported in parentheses. \\
\textsuperscript{b} The acronym stands for fc-CCSD(T)-F12/cc-pVDZ-F12.
}
\end{table}

While the geometry schemes discussed above do not incorporate the CP corrections, four paradigmatic complexes (well representing different types of interactions) have been selected to address (i) the difference between CP and NCP corrected geometries and (ii) the difference between the conventional and explicitly-correlated (F12) CBS+CV schemes. The results are collected in Table \ref{tab:geometriecpncp}.
As far as the comparison between CP and NCP geometries is concerned, the differences are very small, i.e. on the order of few m\AA. This gives support to the choice of defining geometry schemes without CP corrections. The conclusion is that the computational cost due to incorporation of CP corrections is not necessary owing to the extrapolation to the CBS limit performed in all ChS/ChS-F12 approaches. From Table \ref{tab:geometriecpncp}, it is also noted that the differences between conventional and F12 CBS+CV schemes are small as well, these also being on the order of few m\AA. The last comment concerns the fc-CCSD(T)-F12/cc-pVDZ-F12 level. Its performance is nearly as good as that of composite approaches, with deviations of about 0.01 \AA\ for inter-molecular distances. 

In summary, the extrapolation to the CBS limit of the CCSD(T)-F12 geometries is non trivial and the analysis of the convergence of such methods must be carefully checked. The extrapolation formula employed, either $n^{-3}$ or $n^{-5}$, introduces an error of 3 to 5 m\AA, which is close to the error associated with the use of CP or NCP geometries. While F12 basis sets have been purposely tailored for F12 methodologies, it seems that there is no convenience to employ them in comparison with the ``seasonal" jun-cc-pV$n$Z sets. Indeed, for the latter family, triple- and quadruple-zeta basis sets have sizes comparable to those of the double- and triple-zeta sets of the F12 family. However, the``seasonal" basis sets guarantee a solid extrapolation to the CBS limit, also including in the most consistent way diffuse functions. Furthermore, this equivalence between the two families of basis sets is confirmed by the comparison between CCSD(T)-F12/jun-cc-pVTZ (see Table \ref{tab:geomcomplessi_si}) and CCSD(T)-F12/cc-pVDZ-F12 (Table \ref{tab:geometriecpncp}) geometries. These give results with an agreement better than 6 m\AA, with the cc-pVDZ-F12 basis set giving typically shorter bonds.  A final note concerns the full CCSD(T) schemes (conventional and F12) using the aug-cc-pV$n$Z-F12 basis sets (see Table \ref{tab:geomparadigmatic_si}). It is apparent that their performance is very similar to that of junCBS+CV-F12, but at an increased computational cost.

In conclusion, our suggestion is to employ conventional methods within composite schemes to account for the extrapolation to the CBS limit and the CV correction, such as the junChS model, and to employ the F12 explicitly-correlated methods within simple geometry optimizations, possibly improved by the inclusion of the CV term in an additive manner. In view of the incorporation of the CV correction and the use of a larger basis set for the CCSD(T) term, we can consider that the junChS model provides improved results with respect to the CBS-georef counterparts.

\subsubsection{Semi-experimental equilibrium inter-molecular parameters}

\begin{table}[t!]
\centering
\caption{Structural parameters\textsuperscript{a} for the \ce{FA\bond{...}H2O} complex (distances in \AA\ and angles in degrees).}
\label{tab:semiexpformh2o}

\resizebox{\textwidth}{!}{%
    \setlength{\extrarowheight}{2pt}
\begin{tabular}{l c c c | c c c c }
\cline{1-5}
\textbf{\ce{HCONH2}}       &  revDSD &  junChS    &  junChS-F12 & r$_{SE}$\textsuperscript{b}  &   \multicolumn{3}{c}{\multirow{13}{*}{}} \\ \cline{1-5}
$r$(C-H)                   & 1.1028  & 1.0994      &   1.0995     &   1.097(3)   &                     \\ 
$r$(C=O)                   & 1.2144  & 1.2088      &   1.2084     &   1.212(2)   &                    \\ 
$\theta$(NCH)              & 112.56  & 112.69      &   112.73     &   112.4(13)   &                     \\ 
$r$(C-N)                   & 1.3582  & 1.3539      &   1.3533     &   1.354(2)   &                    \\ 
$\theta$(NCO)              & 124.73  & 124.56      &   124.55     &   124.2(5)   &                   \\ 
$r$(NH$_{trans}$)          & 1.0035  & 1.0004      &   1.0003     &   1.017(2)   &                    \\ 
$\theta$(CNH$_{trans}$)    & 121.11  & 121.11      &   121.09     &   120.5(2)   &                   \\ 
$r$(NH$_{cis}$)            & 1.0061  & 1.0030      &   1.0030     &   1.008(2)   &              \\ 
$\theta$(CNH$_{cis}$)      & 119.37  & 119.21      &   119.19     &   119.9(1)   &         \\ \cline{1-5}
\textbf{\ce{H2O}}          & revDSD  &  junChS    &  junChS-F12 & r$_{SE}$\textsuperscript{c}  &         \\ \cline{1-5}                             $r$(H-O)                   & 0.9610  & 0.9563      & 0.9565       & 0.9573(1)    &         \\
$\theta$(HOH)              & 104.46  &  104.49     & 104.55       &   104.53(1)  &         \\ \midrule
                           &         &             &              &                 \multicolumn{4}{c}{r$_{SE}$}                       \\ \cline{5-8}
                           &         &             &              &            &    \multicolumn{3}{c}{TM}                  \\ \cline{6-8}
\textbf{\ce{HCONH2\bond{...}H2O}}   & revDSD  &  junChS    &  junChS-F12  & fixed\textsuperscript{d} &   revDSD  &  junChS         &  junChS-F12          \\ \midrule
$r$(C=O)                   &  1.2246 &     1.2192  &   1.2190     &    -       & 1.2223    &   1.2224        &     1.2226           \\
$r$(N-C)                   &  1.3474 &     1.3432  &   1.3430     &    -       & 1.3431    &   1.3433        &     1.3437           \\
$r$(C-H)                   &  1.1002 &     1.0971  &   1.0970     &    -       & 1.0945    &   1.0947        &     1.0945           \\
$r$(N-H$_{trans}$)         &  1.0035 &     1.0004  &   1.0000     &    -       & 1.0171    &   1.0170        &     1.0167           \\
$r$(N-H$_{cis}$)           &  1.0121 &     1.0094  &   1.0090     &    -       & 1.0020    &   1.0143        &     1.0140           \\
$r$(O$_w$-H$_{cis}$)*      &  2.0522 &     2.0328  &   2.0300     &   2.128(3) & 2.136(3)  &  2.122(3)       &    2.122(3)          \\
$r$(O-H$_c$)               &  0.9746 &     0.9694  &   0.9700     &    -       & 0.9708    &   0.9704        &     0.9708           \\
$r$(O-H$_b$)               &  0.9592 &     0.9547  &   0.9550     &    -       & 0.9554    &   0.9557        &     0.9587           \\
$\theta$(NCO)              &  125.09 &     124.93  &   124.92     &    -       & 124.57    &  124.57         &    124.57            \\
$\theta$(HCN)              &  113.43 &     113.51  &   113.54     &    -       & 113.27    &  113.22         &    113.21            \\
$\theta$(H$_{trans}$NC)    &  120.17 &     120.13  &   120.12     &    -       & 119.56    &  119.52         &    119.52            \\
$\theta$(H$_{cis}$NC)      &  119.66 &     119.53  &   119.54     &    -       & 120.19    &  120.22         &    119.55            \\
$\theta$(O$_{w}$H$_{cis}$N)*&136.10&     135.48  &   136.39     &  133.6(1)  & 133.6(1)  &  134.0(1)       &    134.0(1)          \\
$\theta$(HOH)              &  106.68 &     106.76  &   106.79     &    -       & 106.75    &  106.81         &    106.78            \\
$\theta$(H$_c$O$_{wat}$H$_{cis}$)*&82.44&  83.31   &    83.13     &  71.8(4)   &  69.5(4)  &   69.2(5)       &     69.2(5)         \\ \midrule
$\sigma$                   &         &             &     & 6.4$\times$10$^{-3}$& 6.83$\times$10$^{-3}$ & 7.96$\times$10$^{-3}$ &   7.96$\times$10$^{-3}$      \\ \bottomrule
\end{tabular}
}

\footnotesize
\textsuperscript{a} Inter-molecular parameters are denoted with an asterisks. \\
\textsuperscript{b} Taken from Ref. \citenum{Alessandrini2016}.
\textsuperscript{c} Taken form Ref \citenum{Piccardo2015}. \\
\textsuperscript{d} The parameters not reported are fixed at the values of the corresponding $r_e^{SE}$ values of the isolated monomers.
\end{table}

Owing to the direct connection between rotational constants and molecular structure, microwave spectroscopy can be effectively exploited to obtain accurate geometrical parameters for molecular systems with a non-vanishing dipole moment and sufficiently stable in the gas-phase.\cite{gordy1984microwave} 
While a pure experimental approach cannot be exploited in the majority of cases,\cite{puzzarini2018diving} the so-called semi-experimental (SE) approach\cite{pulay1978cubic} allows the determination of equilibrium structures of experimental quality, the so-called SE equilibrium structures, $r_e^{SE}$. These are obtained from a least-squares fit of the SE equilibrium rotational constants, which are, in turn, derived from the experimental ground-state counterparts by subtracting vibrational corrections. The latter terms can be very effectively computed by hybrid density functionals in conjunction with double-/triple-zeta basis sets (B3LYP-D3(BJ)/SNSD\cite{becke,barone2008development} in the present case).\cite{piccardo2015semi} Whenever experimental information is available for a sufficient number of isotopologues, a complete structural determination is possible, but this is rarely the case for molecular complexes. As a consequence, the application of the SE approach in such cases implies fixing the intra-molecular parameters at those of the isolated fragments and then fitting the most significant inter-molecular parameters.

\begin{table}[t!]
\centering
\caption{Structural parameters\textsuperscript{a} for the \ce{DMS\bond{...}SO2} complex (distances in \AA\ and angles in degrees).}
\label{tab:semiexpSS}

\resizebox{\textwidth}{!}{%
    \setlength{\extrarowheight}{2pt}
\begin{tabular}{l c c c | c c c c}
\cline{1-5}
\textbf{\ce{(CH3)2S}}      &  revDSD &  junChS  &  junChS-F12  & r$_{SE}$\textsuperscript{b}    & \multicolumn{3}{c}{\multirow{14}{*}{}} \\ \cline{1-5}
$r$(C-S)                   & 1.8058  &  1.7984   &      1.7977  &      1.79863(13) &     \\   
$\theta$(CSC)              &  98.69  &  98.43    &      98.49   &      98.58000(81)  &      \\   
$r$(H$_2$-C)               & 1.0900  &  1.0867   &      1.0870  &      1.08857(38) &      \\   
$\theta$(H$_2$CS)          &  107.38 &   107.45  &       107.46 &      107.4196(69) &      \\   
$r$(H$_{1}$-C)             & 1.0910  &  1.0879   &        1.088 &      1.08972(47) &       \\   
$\theta$(H$_{1}$CS)        &  110.87 & 110.75    &       110.79 &      110.688(29) &      \\   
$\phi$(H$_1$CSH$_2$)       & -118.88 & -118.98   &     -118.96  &     -119.053(44) &       \\   
$\phi$(CSXX)               & 130.66  & 130.78    &       130.76 &     129.15  &       \\   
$\phi$(H$_1$CSC)           &  -61.12 & -61.02    &       -61.04 &     -60.95  &     \\   
\cline{1-5}
\textbf{\ce{SO2}}          &  revDSD &  junChS  &  junChS-F12  & r$_{SE}$\textsuperscript{c}    & \multicolumn{3}{c}{\multirow{14}{*}{}} \\ \cline{1-5}
$r$(S-O)                   &  1.4421 &  1.4298   &   1.4288     &    1.4307858(15)  &         \\   
$\theta$(OSO)              &  119.28 &  119.29   &   119.24     &    119.329872(81)   &         \\   
                           &         &           &              &         \multicolumn{4}{c}{r$_{SE}$}        \\ \cline{5-8}
                           &         &           &              &            &   \multicolumn{3}{c}{TM}  \\ \cline{6-8}
\textbf{\ce{(CH3)2S\bond{...}SO2}}  & revDSD  &  junChS  &  junChS-F12 & fixed\textsuperscript{d}  &   revDSD   &  junChS     &  junChS-F12  \\ \midrule
$r$(S1-S2)*              &  2.9288     & 2.9570    &     2.9632   & 2.944(2)   &   2.943(3) &    2.945(2) &   2.944(2)   \\
$r$(S1-O)               &  1.4500     & 1.4266    &     1.4358   &   -        &   1.4387   &    1.4276   &   1.4378     \\
$\theta$(OS1S2)*         &  94.92      & 94.18     &     94.21    &  95.5(1)   &   95.4(1)  &  95.1(1)    &   95.4(1)    \\
$\phi$(O-S1-S2-X)        &  -121.07    &  -120.76  &     -121.08  &   -        &   -121.01  &  -120.11    &  -120.81     \\
$r$(C-S2)               &  1.8029     & 1.7928    &      1.7958  &   -        &   1.7957   &    1.7930   &   1.7968     \\
$\theta$(CS2S1)*         &  91.52      & 91.16     &      91.1    &  91.5(1)   &   91.6(1)  &  91.6(1)    &   91.6(1)    \\
$\phi$(C-S2-S1-X)*       &  130.17     &  130.30   &    130.32    &  130.7(3)  &   130.5(3) &  130.7(3)   &   130.6(3)   \\
$r$(H1-C)           &  1.0912     & 1.0886    &     1.0887   &    -       &   1.0898   &    1.0904   &    1.0904    \\
$\theta$(H1CS3)         &  110.60     &  110.66   &      110.63  &    -       &   110.42   &  110.60     &    110.53    \\
$\phi$(H1CS3C)          &  -63.89     &  -64.34   &     -64.00   &    -       &   -63.72   &  -64.26     &   -63.90     \\
r(H2-C)             &  1.0895     &  1.0866   &      1.0865  &    -       &   1.0881   &    1.0885   &  1.0881      \\
$\theta$(H2CS3)      &  107.25     &  107.35   &      107.35  &    -       &   107.29   &  107.32     &  107.31      \\
$\phi$(H2CS2H1)      &  -118.69    & -118.74   &   -118.73    &    -       &   -118.85  &  -118.81    &  -118.82     \\
$r$(H3C)            &  1.0901     &  1.0873   &      1.0872  &    -       &   1.0887   &    1.0891   &  1.0889      \\
$\theta$(H3CS2)      &  109.93     &  109.94   &     109.94   &    -       &   109.75   &  109.89     &  109.84      \\
$\phi$(H3CS2C)       &  58.563     &  58.05    &    58.36     &    -       &   58.39    &  57.98      &  58.27       \\ \midrule
$\sigma$               &             &           &      & 4.8 $\times$ 10$^{-3}$ &6.3 $\times$ 10$^{-3}$ & 4.4 $\times$ 10$^{-3}$ & 5.1 $\times$ 10$^{-3}$   \\ \bottomrule
\end{tabular}
}

\footnote

\textsuperscript{a} Inter-molecular parameters are denoted with an asterisks. \\
\textsuperscript{b} Taken from Ref. \citenum{DEMAISON2010229}.
\textsuperscript{c} Taken from Ref \citenum{Demaison2021} \\
\textsuperscript{d} The parameters not reported are fixed at the values of the corresponding $r_e^{SE}$ values of the isolated monomers. 
\end{table}

\begin{figure}[t!]
    \centering
\includegraphics[width=12cm]{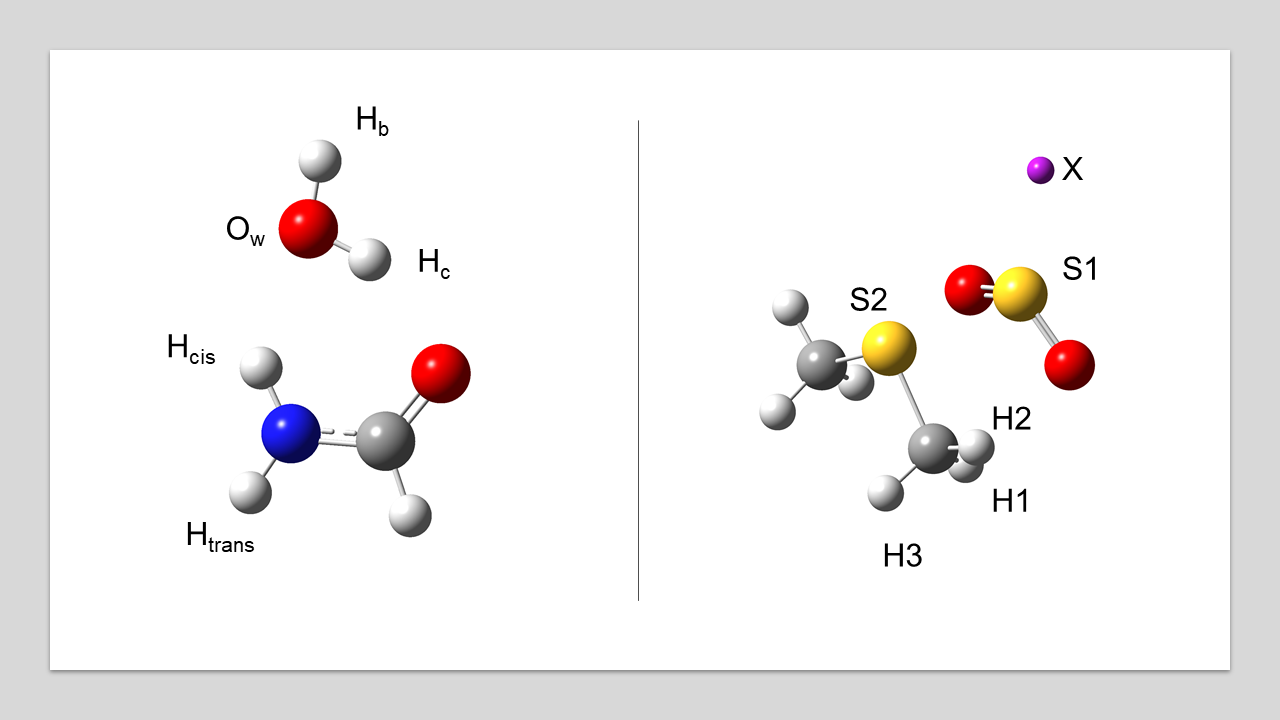}
    \caption{\ce{FA\bond{...}H2O} and \ce{DMS\bond{...}SO2} complexes: structure and atom labeling.}
    \label{Atomlabeling}
\end{figure}

In the framework of this work, it is worthwhile testing whether ChS/ChS-F12 models are able to deliver accurate geometrical parameters to be employed in the determination of $r_e^{SE}$'s of molecular complexes. 
To this end, we have selected two representative complexes, namely formamide-water (\ce{FA\bond{...}H2O}) and \ce{DMS\bond{...}SO2}.
Let us start from the structures of the isolated molecules. For all of them, $r_e^{SE}$ determinations are feasible, the results being collected in Tables \ref{tab:semiexpformh2o} and \ref{tab:semiexpSS} for \ce{FA\bond{...}H2O} and \ce{DMS\bond{...}SO2}, respectively, with the atom labeling being provided in Figure~\ref{Atomlabeling}. In these tables, the optimized geometries at the revDSD, junChS and junChS-F12 levels are also reported. From the inspection of Tables \ref{tab:semiexpformh2o} and \ref{tab:semiexpSS}, it is evident that the junChS and junChS-F12 models provide results very similar to the $r_e^{SE}$ counterparts, with discrepancies of a few m\AA ~ for bond lengths and a few tenths of degree for valence angles. The close agreement between the conventional and explicitly-correlated versions of the junChS composite scheme is also retained for the inter-molecular parameters of both the studied complexes, with however increased absolute deviations from the $r_e^{SE}$. In addition, these tables further confirm the reliability of the revDSD model.

Before further proceeding with the discussion of the SE equilibrium structures, it is necessary to explain how they have been determined. As mentioned above, when there is a lack of experimental data, the intra-molecular parameters can be fixed to those of the isolated monomers. This is the case of the results of Tables \ref{tab:semiexpformh2o} and \ref{tab:semiexpSS} denoted ``fixed", with the $r_e^{SE}$ values of the monomers being employed. In the other cases, we resorted to the template molecule (TM) approach\cite{piccardo2015semi} to evaluate the parameter values to be kept fixed ($r_e(\text{fixed})$):
\begin{equation}
r_e(\text{fixed}) = r_e^{level} + \Delta r_e(\text{TM})
\label{TM}
\end{equation}
where $r_e^{level}$ is the value of a generic intra-molecular parameter of the molecular complex evaluated at a given level of theory (revDSD, junChS or junChS-F12, in the present case) and $\Delta r_e(\text{TM})$ is correction to be applied based on the corresponding value within the isolated monomer:
\begin{equation}
\Delta r_e(\text{TM}) = r_e^{SE} - r_e^{level} 
\label{CORR}
\end{equation}

Focusing on the various SE equilibrium structures reported in Tables \ref{tab:semiexpformh2o} and \ref{tab:semiexpSS}, it is interesting to note that using the revDSD, junChS or junChS-F12 level in the TM approach leads to very similar results. This suggests that, whenever exploiting this methodology for accurate structural determinations of molecular complexes, there is no need of composite schemes, but instead a very cost-effective level of theory such as revDSD is largely sufficient.

\subsection{The junChS and junChS-F12 databases}

The previous sections have demonstrated the reliability and the adequate accuracy of the revDSD optimized geometries as reference structures, which have thus been retained for setting up a database of accurate interaction energies. The CP, NCP and half-CP interaction energies for the A14 complexes using both the junChS and junChS-F12 schemes are collected in Table \ref{tab:confrontojuncomplessi}. While the statistics for these two approaches have already been discussed, from this table it is apparent that, in absolute terms, the differences between the two models are only fractions of \SI{}{\kilo\joule\per\mol}. It is interesting to note that CP and NCP results are always very similar, with differences in almost all cases well within 0.1 \SI{}{\kilo\joule\per\mol}, thus pointing out the effectiveness of the extrapolation to the CBS limit in recovering the BSSE. However, the CP-NCP differences are systematically smaller for the junChS-F12 scheme.

\begin{table}[t!]
\centering
\caption{Interaction energies of the A14 complexes with revDSD reference geometries. Energies in \SI{}{\kilo\joule\per\mol}. }
\label{tab:confrontojuncomplessi}
\begin{tabular}{lcccccc}
\toprule
          & \multicolumn{3}{c}{junChS} & \multicolumn{3}{c}{junChS-F12} \\ \midrule
                        & CP     & NCP    & half-CP& CP     & NCP    & half-CP  \\ \cmidrule(l){2-7} 
\ce{H2O\bond{...}H2O}   & -21.10 & -21.37 & -21.24 & -21.00 & -21.11 & -21.05 \\
\ce{NH3\bond{...}NH3}   & -13.30 & -13.34 & -13.32 & -13.20 & -13.19 & -13.20 \\
\ce{HF\bond{...}HF}     & -19.45 & -19.59 & -19.52 & -19.25 & -19.41 & -19.33 \\
\ce{HCN\bond{...}HCN}   & -19.88 & -19.70 & -19.79 & -19.84 & -19.99 & -19.91 \\
\ce{CH4\bond{...}CH4}   & -2.25  & -2.22  & -2.24  & -2.20  & -2.16  & -2.18  \\
\ce{CH2O\bond{...}CH2O} & -19.23 & -19.43 & -19.33 & -18.96 & -19.10 & -19.03 \\
\ce{C2H4\bond{...}C2H4} & -4.75  & -4.78  & -4.77  & -4.66  & -4.70  & -4.68   \\
\ce{H2O\bond{...}C2H4}  & -10.86 & -11.01 & -10.93 & -10.77 & -10.80 & -10.79  \\
\ce{H2O\bond{...}CH4}   & -2.78  & -2.79  & -2.79  & -2.77  & -2.75  & -2.76   \\
\ce{H2O\bond{...}NH3}   & -27.57 & -27.69 & -27.63 & -27.47 & -27.52 & -27.50  \\
\ce{NH3\bond{...}CH4}   & -3.24  & -3.24  & -3.24  & -3.23  & -3.22  & -3.22   \\
\ce{NH3\bond{...}C2H4}  & -5.89  & -5.99  & -5.94  & -5.82  & -5.84  & -5.83   \\
\ce{HF\bond{...}CH4}    & -7.13  & -7.14  & -7.13  & -7.04  & -7.06  & -7.05   \\
\ce{C2H4\bond{...}CH2O} & -6.94  & -7.07  & -7.01  & -6.83  & -6.91  & -6.87   \\ \bottomrule
\end{tabular}%
\end{table}

To extend the representativeness of both our benchmark study and the database to be set up, we have selected some molecules including third-row atoms either taken from our previous work\cite{alessandrini2020} or purposely selected for this study (B9 dataset). Furthermore, some additional larger molecular complexes (C6 dataset) have been chosen, most of them from the S66 dataset,\cite{s66} in order to improve the coverage for the H, C, N, O atoms\cite{s66}. The results are collected in Table~\ref{tab:energietestcase}, where the junChS and junChS-F12 models  (both employing revDSD reference geometries) are compared, with CP, NCP, and half-CP interaction energies being considered. In Table~\ref{tab:energietestcase}, previous available data are also provided. In this respect, it should be noted that the junChS results for \ce{FH2P\bond{...}H2S}, \ce{FH2P\bond{...}NH3} and \ce{DMS\bond{...}SO2} were already reported in ref. \citenum{alessandrini2020}; however, small discrepancies ($<$0.1 \SI{}{\kilo\joule\per\mol}) can be noted due to the use of a different reference geometry (B2PLYP-D3/maug-cc-pVTZ in ref. \citenum{alessandrini2020}). The situation is different for the complexes taken from the S66 dataset, for which differences as large as 3 \SI{}{\kilo\joule\per\mol} can be observed. 
Actually, this is somewhat expected, the level of theory employed in the S66 dataset being less accurate than our junChS and junChS-F12 approaches. Indeed, in ref. \citenum{s66}, starting from HF/aug-cc-pVQZ energies, the CBS contribution is evaluated at the MP2 level using the aug-cc-pVTZ and aug-cc-pVQZ basis sets, and the effect of triple excitations is incorporated via the CCSD(T)-MP2 energy difference computed using the aug-cc-pVDZ basis set. Furthermore, the CV correction is not included.
The cyclopentene-water complex was previously investigated at the ChS level,\cite{cyclopentene} thus the improvement is due to the balanced incorporation of the diffuse function effects.

\begin{table}[t!]
\centering
\caption{junChS and junChS-F12 interaction energies (revDSD reference geometry) for the B9 and C6 datasets. Values in \SI{}{\kilo\joule\per\mol}.}
\label{tab:energietestcase}
\resizebox{\textwidth}{!}{%
\begin{tabular}{lccccccc}
\toprule
                & \multicolumn{3}{c}{junChS} & \multicolumn{3}{c}{junChS-F12} \\ \midrule
                & CP      & NCP    & half-CP & CP       & NCP      & half-CP & Literature \\ \cmidrule(l){2-7} 
B9 dataset \\   \cline{1-1}              
\ce{FH2P\bond{...}H2S}         & -14.92 & -15.42 & -15.17 & -14.80 & -14.89 & -14.85 & -14.94$^a$ \\
\ce{FH2P\bond{...}NH3}         & -29.46 & -30.03 & -29.74 & -29.08 & -29.24 & -29.16 & -29.54$^a$ \\
\ce{H2O\bond{...}H2S}          & -12.18 & -12.41 & -12.30 & -12.23 & -12.25 & -12.24 &          \\
\ce{H2O\bond{...}PH3}          & -10.77 & -10.95 & -10.86 & -10.76 & -10.77 & -10.76 &          \\
\ce{CH3NH2\bond{...}HCl}       & -52.27 & -52.81 & -52.54 & -52.32 & -52.41 & -52.37 &          \\
\ce{OCS\bond{...}CH4}          & -4.36  & -4.45  & -4.41  & -4.25  & -4.29  & -4.27  &          \\
\ce{OCS\bond{...}H2O}          & -7.96  & -8.09  & -8.02  & -7.87  & -7.89  & -7.88  &          \\
\ce{SO2\bond{...}H2S}          & -12.22 & -12.45 & -12.34 & -12.11 & -12.20 & -12.15 &          \\
\ce{DMS\bond{...}SO2}          & -33.53 & -34.40 & -33.97 & -32.97 & -33.36 & -33.17 & -34.04$^a$ \\  \cmidrule(l){2-7}
C6 dataset \\  \cline{1-1} 
\ce{Cyclopentene\bond{...}H2O} & -16.59 & -16.71 & -16.65 & -16.37 & -16.44 & -16.40 &  -12.8$^b$ \\
\ce{H2O\bond{...}Peptide}      & -35.16 & -35.54 & -35.35 & -34.91 & -35.05 & -34.98 & -33.89$^c$; -32.31$^d$\\
\ce{CH3NH2\bond{...}CH3NH2}    & -17.82 & -17.94 & -17.88 & -17.64 & -17.67 & -17.66 & -17.41$^c$; -15.23$^d$ \\
\ce{CH3NH2\bond{...}Pyridine}  & -20.17 & -20.27 & -20.22 & -19.94 & -19.99 & -19.96 & -16.61$^c$; -13.19$^d$ \\
\ce{CH3OH\bond{...}Pyridine}   & -28.18 & -28.65 & -28.42 & -28.08 & -28.21 & -28.15 & -31.00$^c$; -28.57$^d$ \\
\ce{Pyridine\bond{...}Pyridine}& -16.32 & -16.58 & -16.45 & -16.12 & -16.26 & -16.19 & -16.32$^c$; -9.97$^d$ \\  \bottomrule
\end{tabular}%
}

\footnotesize
\textsuperscript{a} Ref. \citenum{alessandrini2020}: at the junChS level \\
\textsuperscript{b} Vibrational ground-state dissociation energy taken from Ref. \citenum{cyclopentene}: ChS CP-corrected electronic energy augmented by harmonic zero-point energy at the B2PLYP-D3/maug-cc-pVTZ-$d$H level.  \\
\textsuperscript{c} Ref. \citenum{s66}: see text for the description of the level of theory.\\ 
\textsuperscript{d} Ref. \citenum{Tew2014}: CCSD/CBS energy computed as the sum of HF/aug-cc-pVQZ energy and PNO-CCSD-F12/aug-cc-pVTZ energy.
\end{table}

Focusing on the comparison between the junChS and junChS-F12 schemes, inspection of Table~\ref{tab:energietestcase} shows that the two approaches provide very similar results, with a maximum difference of 0.6 \SI{}{\kilo\joule\per\mol} for CP-corrected interaction energies and 0.8 \SI{}{\kilo\joule\per\mol} for the NCP counterparts. Furthermore, it is noted that, in absolute terms, the CP-corrected junChS interaction energies are usually larger than the junChS-F12 counterparts by about 0.1-0.2 \SI{}{\kilo\joule\per\mol}. As expected in view of the extrapolation to the CBS limit, the CP corrections are small. As noted for the A14 dataset, the CP-NCP energy differences are slightly more pronounced for schemes involving conventional methods. Indeed, the CP correction to junChS-F12 interaction energies is usually as low as 0.1 \SI{}{\kilo\joule\per\mol}, the exception being the DMS-\ce{SO2} complex, for which it is about 0.4 \SI{}{\kilo\joule\per\mol}. This suggests that the CP correction can be safely neglected when employing the junChS-F12 model, with a remarkable saving of computer time. The average CP correction, $\sim$0.3 \SI{}{\kilo\joule\per\mol}, is quite small also for the conventional junChS model. However, it is comparable (if not larger) than the expected error for this approach; therefore, its neglect is not recommended. The CP correction of the interaction energy of DMS-\ce{SO2} is particularly large, i.e. 0.9 \SI{}{\kilo\joule\per\mol}, also when considering the conventional model. 

Another remarkable aspect of the results for the B9 dataset is that the trends for molecules containing third-row atoms  are extremely similar to those only bearing first-/second-row elements, with the only exception being the DMS-\ce{SO2} complex. For the systems of Table \ref{tab:energietestcase} previously investigated,\cite{s66,alessandrini2020} since the junChS-F12 model does not change the trends in any significant manner, we only recall that the present results are the most accurate currently available for both the B9 and C6 dataset.

In conclusion, based on the results of this study and those of ref.~\citenum{alessandrini2020}, the junChS and junChS-F12 models allowed us to establish a database of accurate interaction energies at a limited computational cost, the actual version of this database consisting of the A14, B9 and C6 datasets. The present work has demonstrated the applicability of the junChS and junChS-F12 approaches to molecular complexes involving third-row atoms without any degradation in the accuracy as well as the possibility of its application to large systems. As a part of the  database, the structural parameters of the A14 dataset are also provided. According to the discussion of a previous section, only the junChS and CCSD(T)-F12/jun-cc-pVTZ+CV(MP2/cc-pwCVTZ) levels have been retained. The database will be extended in order to incorporate more information for both interaction energies and equilibrium structures.

\section{Concluding remarks}

Composite schemes based on explicitly-correlated (F12) approaches for the description of non-covalent complexes have been defined starting from the well-tested ``cheap" methodology for conventional methods. The performance has been analyzed in detail for systems ruled by different inter-molecular interactions employing a representative dataset including atoms belonging to the first two rows of the periodic table. While these composite schemes have been first introduced for the accurate evaluation of interaction energies, they have then been extended to structural determinations.

Among the different approaches investigated, the so-called junChS-F12 model together with its conventional counterpart, junChS, and a modified version obtained by removing $d$ and $f$ polarization functions on hydrogen atoms, the jun-($d,f$)H\_ChS-F12 model, have been found to be the best performing schemes. The junChS-F12 and jun-($d,f$)H\_ChS-F12 approaches, tested on the dataset denoted as A14, showed average relative errors of 0.68\% and 0.65\% (for CP-corrected interaction energies), respectively, which means --in absolute terms-- average deviations of 0.05 \SI{}{\kilo\joule\per\mol}.
Interestingly, if the CP correction is neglected, thus further reducing the computational cost, the performances of the different ``cheap" variants show a limited worsening, with the average relative errors of junChS-F12 and jun-($d,f$)H\_ChS-F12 models increasing to 1.10\% and 1.07\%, respectively.

The present study confirms the outcomes of our previous work that led to the definition of the junChS scheme\cite{alessandrini2020}: (i) the inclusion of up to $d$ diffuse functions on non-hydrogen atoms is mandatory in the CCSD(T) computational step for both conventional and explicitly-correlated approaches. For this purpose, the aug-cc-pVDZ-F12 and jun-cc-pVTZ basis sets (which, as explained in the text, have comparable dimensions) are the smallest options for accurate results and, actually, deliver comparable results; (ii) the extrapolation to the CBS limit at the MP2(-F12) level improves significantly the accuracy of the results without any relevant increase of the computational cost for both conventional and explicitly-correlated approaches; (iii) the CV correlation correction always represents a quite small contribution that cannot, however, be neglected in view of the very small absolute errors. 

The junChS-F12 approach slightly outperforms its conventional junChS counterpart, the latter showing a relative error of 1.38\%. For quite small systems, single-point junChS(-F12) computations require no more than twice the time of the underlying coupled-cluster step and are one order of magnitude faster than the CBS+CV counterparts. Enlarging the dimension of the systems increases the effectiveness of the junChS(-F12) model because of the favorable scaling of MP2(-F12) computations with respect to CCSD(T)(-F12), with the performance of MP2(-F12) being further enhanced by the resolution of identity and other acceleration techniques. Concerning the coupled-cluster step, inclusion of explicit correlation increases the computer time by 20\% at most, a burden which is accompanied by a non negligible improvement in accuracy. Replacing MP2-F12 with CCSD(T)-F12 in the extrapolation step (thus leading to the junCBS+CV-F12 model) increases the computational cost by at least one order of magnitude without improving the performance of the composite scheme. Indeed, for CP-corrected interaction energies, the average relative error increases from 0.68\% to 0.79\% and the average absolute error from  \SI{0.05}{\kilo\joule\per\mol} to \SI{0.07}{\kilo\joule\per\mol}. 

Moving to the structural investigation, an important result is that the geometries optimized employing the recent rev-DSD-PBEP86 double-hybrid functional augmented by D3(BJ) empirical dispersion correction and in conjunction with the jun-cc-pVTZ basis set show a reasonably good accuracy in most situations. If improved accuracy is sought, the most effective option is to add the core-valence correlation correction, evaluated at the MP2-F12 level, to fc-CCSD(T)-F12/jun-cc-pVTZ geometries without accounting for the extrapolation to the CBS limit and the basis set superposition error. Indeed, F12 explicitly-correlated methods deliver geometrical parameters close to the basis-set convergence, thus leading to negligible basis set superposition errors.

Although the study of larger systems requires further developments, possibly related to the use of local-correlation treatments (e.g. PNO or DLPNO possibly including explicit correlation\cite{Schmitz2014,dlpno-2021}), the different variants of the so-called ``cheap" composite scheme described in the present paper already represent reliable and effective tools for the investigation of inter-molecular interactions between biomolecule building blocks. In particular, both junChS and junChS-F12 models have been successfully applied to the C6 dataset, which involves rather large systems such as the pyridine dimer.

\begin{acknowledgement}
This work has been supported by MIUR 
(Grant Number 2017A4XRCA) and by the University of Bologna (RFO funds). The SMART@SNS Laboratory (\url{http://smart.sns.it}) is acknowledged for providing high-performance computing facilities.
\end{acknowledgement}

\begin{suppinfo}
In the Supplementary Information the statistics for the following composite schemes applied on top of A14 dataset are reported: junChS-F12 NCP energies (Table \ref{tab:ncpjChSF12_si}), jun-($d,f$)H\_ChS-F12 NCP energies (Table \ref{tab:ncpjChSF12df_si}), junCBS+CV-F12 NCP energies (Table \ref{tab:ncpjCBSF12_si}), and junChS CP and NCP energies (Tables \ref{tab:jchs_cp_si} and \ref{tab:jchs_ncp_si}, respectively). Table \ref{tab:geomcomplessi_si} reports a full account on the geometrical parameters obtained for the A14 dataset for all the computational approaches considered, while Table \ref{tab:cbsgeomparam_si} reports a detailed analysis of the CBS and CV terms for the ChS, junChS and junChS-F12 models. Finally, Table \ref{tab:geomparadigmatic_si} reports CP and NCP geometries for the \ce{H2O\bond{...}H2S}, \ce{H2O\bond{...}H2O}, \ce{HCN\bond{...}HCN} and \ce{CH4\bond{...}CH4} complexes, obtained using the augF12CBS+CV and augF12CBS+CV-F12 composite approaches.
\end{suppinfo}

\bibliography{main}
\includepdf[pages=-]{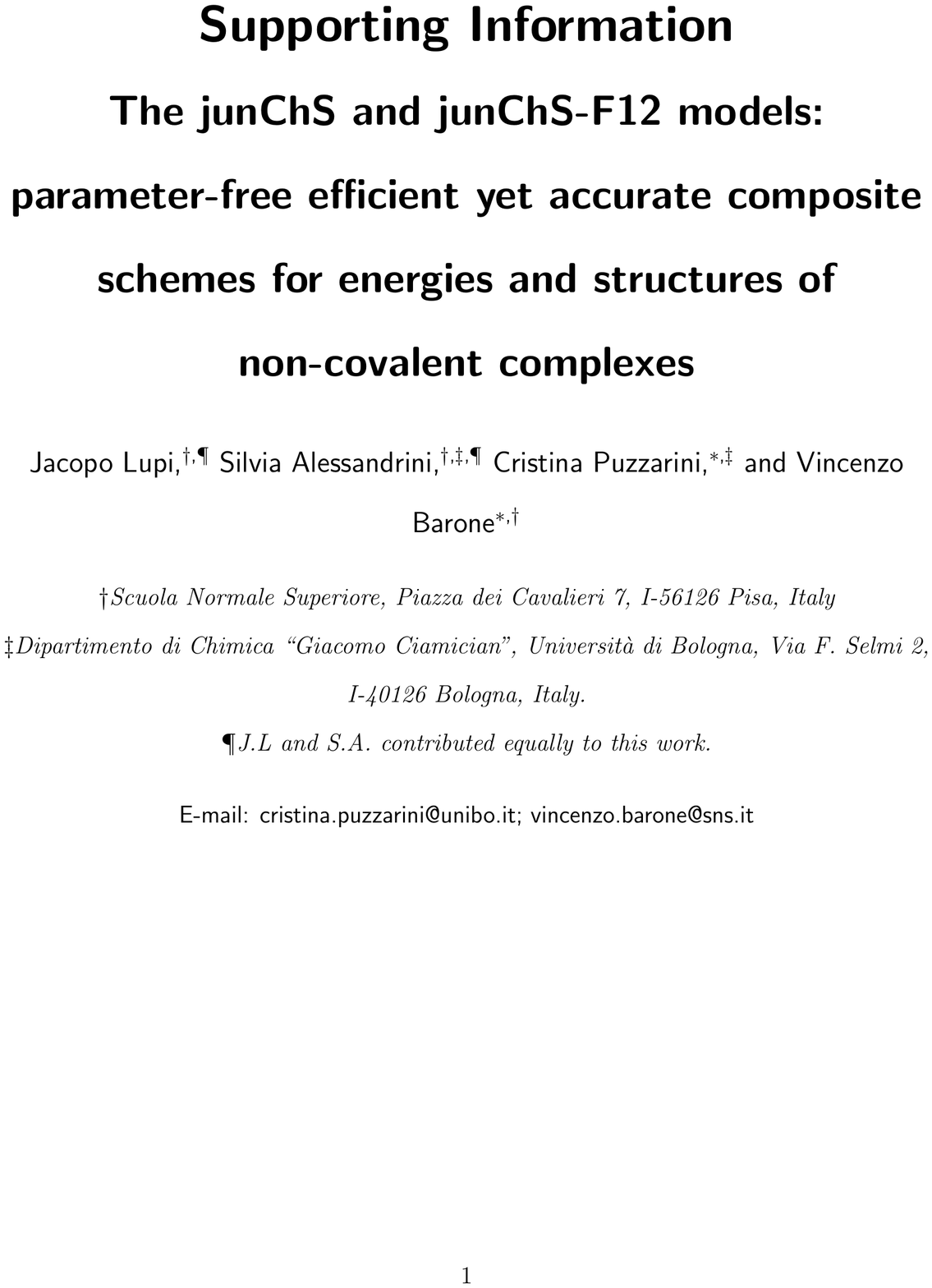}

\end{document}